\documentclass[twocolumn]{article}

\usepackage{imr}
\usepackage{graphicx}

\usepackage{geometry}
\usepackage{subfigure,pgfplots,epsf,graphicx,color}
\usepackage{amsmath,amsfonts,amsthm,multirow}

\usepackage{comment}
\usepackage{rotating}
\usepackage{tabularx}
\usepackage{float}

\usepackage{pdfpages}
\usepackage{afterpage}
\usepackage{wrapfig}
\usepackage{microtype}
\usepackage{fullpage}
\usepackage{bm} 
\usepackage{bbm} 
\usepackage{caption} 

\usepackage{setspace}
\usepackage{lipsum}
\newcommand{\verbatimfont}[1]{\renewcommand{\verbatim@font}{\ttfamily#1}}
\usepackage{graphicx}
\usepackage{epstopdf}
\usepackage{alltt}
\usepackage{color}
\usepackage{multirow}

\usepackage{fancyvrb}
\fvset{fontsize=\scriptsize}
\RecustomVerbatimEnvironment{verbatim}{Verbatim}{}

\usepackage[colorlinks=true]{hyperref}
\hypersetup{
  colorlinks  = true, 
  urlcolor    = blue, 
  linkcolor   = blue, 
  citecolor   = blue, 
}

\DeclareGraphicsExtensions{.gif,.png,.jpg,.pdf,.tiff,.eps}

\newcommand{\dt}{\dftl{t}}

\newcommand{\cE}{\mathcal{E}}

\newcommand{\cN}{\mathcal{N}}

\newcommand{\cP}{\mathcal{P}}

\newcommand{\be}{\begin{equation}}
\newcommand{\ee}{\end{equation}}
\newcommand{\bes}{\begin{equation*}}
\newcommand{\ees}{\end{equation*}}
\newcommand{\bse}{\begin{subequations}}
\newcommand{\ese}{\end{subequations}}

\newcommand{\ux}{u^1}

\def\dO{\partial \Omega}

\def\ee{{\hat {\underline e}}}

\def\dt{ \Delta t }

\def\cN{{\cal N}}
\def\cE{{\cal E}}

\def\scriptO{{{\it O}\kern -.42em {\it `}\kern + .20em}}
\def\RR{{{\rm l}\kern - .15em {\rm R} }}
\def\PP{{{\rm l}\kern - .15em {\rm P} }}
\def\L2{{{\sf L}^2}}
\def\H1{{{\sf H}^1}}
\def\PN2{{\PP_{N}-\PP_{N-2}}}

\def\complex{{{\rm C} \kern - .53em {\rm l} \kern + .38em}}

\def\a1{{ | \lambda_{\min} |}}

\def\l1{{   \lambda_{\min}  }}

\def\bhn{{\hat   {\bf n}}}

\def\bu0{{\underline {\bf 0}}}

\def\br{{\bf r}}

\def\be{{\bf e}}
\def\bu{{\bf u}}
\def\bp{{\bf p}}

\def\bx{{\bf x}}

\def\Oh{{\hat \Omega}}

\def\ug{{\underline g}}

\def\up{{\underline p}}

\def\uv{{\underline v}}

\def\ux{{\underline x}}

\def\u0{{\underline 0}}

\def\ubx{{\bf \ux}}

\def\bbone{\ensuremath{\mathbbm{1}}}

\newcommand{\pp}[2]{\frac{\partial #1}{\partial #2} }

\begin{document}

  \title{{All-Hex Meshing Strategies for Densely Packed Spheres}}

 \author{Yu-Hsiang Lan$^1$ \and Paul Fischer$^{1,2,3}$ \and Elia Merzari$^4$ \and Misun Min$^1$}
 \date{
 $^1$Mathematics and Computer Science Division, Argonne National Laboratory, Lemont, IL, U.S.A.\\
 $^2$Department of Computer Science, University of Illinois at Urbana-Champaign, 
     Urbana, IL, U.S.A.\footnote{Corresponding author: fischerp@illinois.edu}\\
 $^3$Department of Mechanical Science and Engineering, University of Illinois, Urbana, IL, U.S.A.\\
 $^4${Department of Nuclear Engineering, Penn State, University Park, PA, U.S.A.
\vspace*{-.9in}
}
}

  \abstract{
We develop an all-hex meshing strategy for the interstitial space in beds of
densely packed spheres that is tailored to turbulent flow simulations based on
the spectral element method (SEM).  The SEM achieves resolution through elevated
polynomial order $N$ and requires two to three orders of magnitude fewer
elements than standard finite element approaches do.  These reduced element
counts place stringent requirements on mesh quality and conformity.  Our
meshing algorithm is based on a Voronoi decomposition of the sphere centers.
Facets of the Voronoi cells are tessellated into quads that are swept to the
sphere surface to generate a high-quality base mesh.  Refinements to the
algorithm include edge collapse to remove slivers, node insertion to balance
resolution, localized refinement in the radial direction about each sphere, and
mesh optimization.  We demonstrate geometries with $10^2$--$10^5$ spheres using
$\approx$ 300 elements per sphere (for three radial layers), along with mesh
quality metrics, timings, flow simulations, and solver performance.  }

  \keywords{all-hex meshing,
            spectral elements, 
            smoothing, 
            projection, 
            pebble bed reactor}

  \maketitle
  \thispagestyle{empty}
  \pagestyle{empty}

\section{Introduction}

\begin{figure}[t]
\centering
{\setlength{\unitlength}{1.0in}
   \begin{picture}(3.300,3.00)(0,-.1)
      \put(0.40,0.00){\includegraphics[width=2.2in]{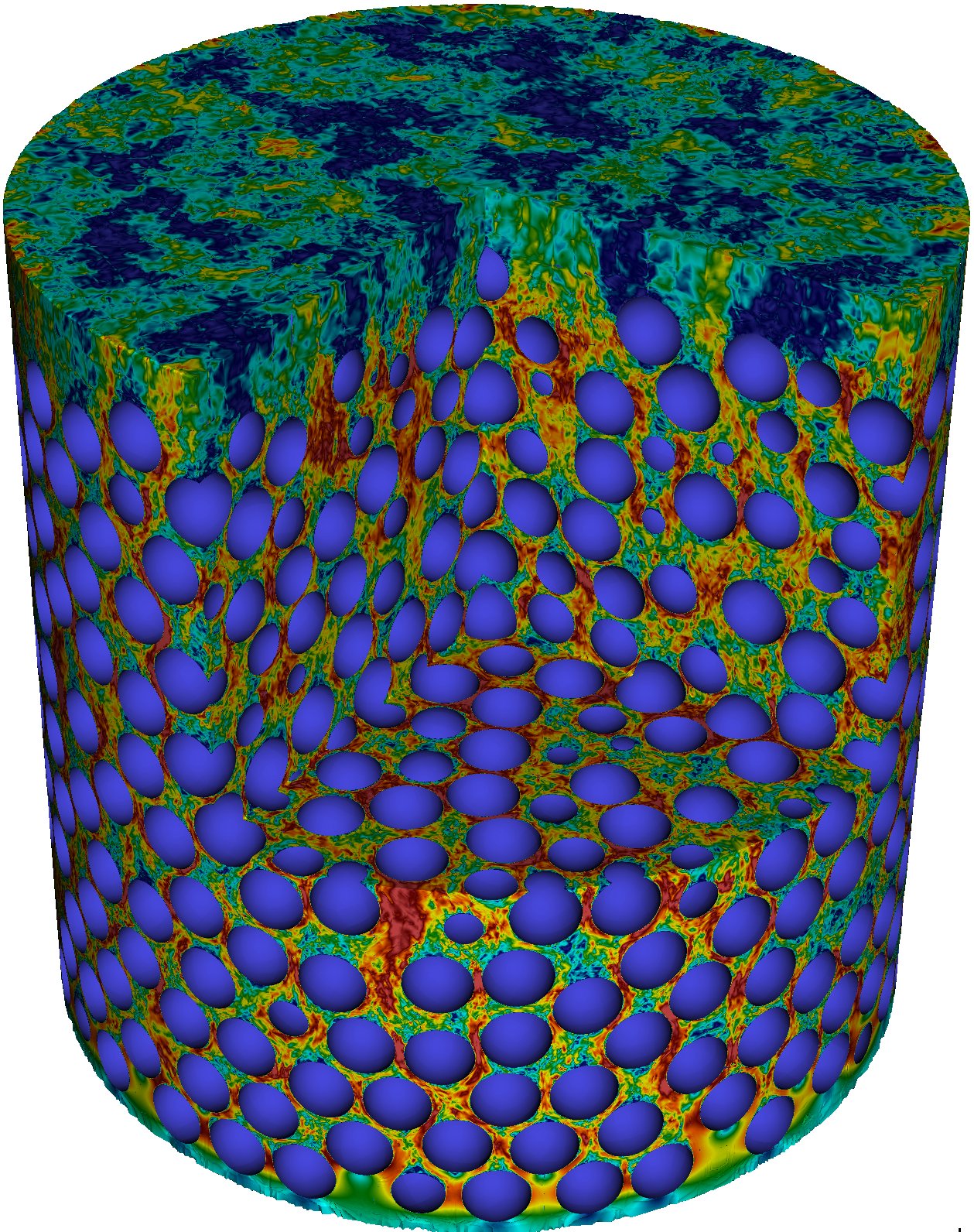}}
   \end{picture}}
\caption{
Example of 
turbulent flow in a packed bed with $\cN=1568$ unit-radius spheres
meshed with $E=524,386$ spectral elements of order $N=7$.
\label{fig:1568}
}
\end{figure}

We are interested in simulating turbulent flow through randomly packed
spherical beds such as illustrated in Fig. \ref{fig:1568}.
Spherical beds are common in many industrial processes in chemical
engineering \cite{packedbed06}. The flow of coolant through packed beds is
of particular interest in the design of pebble-bed reactors, and
researchers  have expressed significant interest in detailed simulations tha
can provide insight into heat transfer in new pebble-bed
designs  \cite{merzari2020a}.
For simulations that resolve turbulent eddies in the flow, high-order
discretizations having minimal numerical dissipation and dispersion provide
high accuracy with a relatively small number of gridpoints, $n$.
The spectral element method (SEM)~\cite{pat84},
which uses local tensor-product bases on curved hexahedral elements, 
is efficient, with memory costs scaling as $O(n)$,
independent of local approximation order $N$ (for $n$ fixed), and 
computational cost that is only $O(nN)$.  Here, $n \approx EN^3$ is the
total number of gridpoints for a mesh comprising $E$ elements.
In contrast, $p$-type finite element methods 
exhibit $O(EN^6)=O(nN^3)$ storage and work complexities,
which effectively limit approximation orders to $N \leq$4.  

For a given resolution $n$, the use of high-order elements with $N=7$--15
implies a 300- to 3000-fold reduction in the number of elements required
when compared with linear finite elements.  The SEM meshing task is thus a 
challenge. We require {\em high-quality meshes with relatively few elements.}  
By contrast, tet- or hex-meshes for linear elements are sufficiently
fine-grained that one can fairly easily repair connections where needed.
Paving/plastering is one all-hex example that illustrates this approach
\cite{stephenson1992,cass1996}.
For the dense-packed sphere problem, however, the distance between the sphere
boundaries and the center of the voids is not large---paved surfaces will
quickly collide, and a large number of elements will be required to
conformally merge the advancing fronts (e.g., \cite{geode}).

An alternative all-hex strategy is to tessellate the void space with
tets and to then convert each tet to four hexes (e.g., \cite{mcdill2004}).
In addition to providing a straightforward path to an all-hex mesh, this
approach leads to fairly isotropic elements that result in reasonable iteration
counts for the pressure Poisson solve, which dominates the cost of most
incompressible flow simulations.  The tet-to-hex strategy has recently been 
pursued by Yuan et al. for packed beds \cite{yuan2020}.   Unfortunately,
the element counts are high. The authors found that they could
only use $N=4$ for the target resolution ($n$) in their simulations, 
which is suboptimal for the SEM where $N>5$ is preferred \cite{fischer20a}.

Here, we propose a meshing algorithm that is based on a Voronoi decomposition
of the sphere centers.  Facets of the Voronoi cells are tessellated into quads
that are swept to the sphere surface to generate a high-quality base mesh.
Refinements to the algorithm include placement of ghost spheres to generate
boundary cells, edge collapse to remove slivers, node insertion to balance
resolution, localized refinement in the radial direction about each sphere, and
mesh optimization.  


\begin{figure} \centering {\setlength{\unitlength}{0.88in}
\begin{picture}(2.300,8.90)(0,0.3)
\put(-.55,0.5){\includegraphics[width=3.33in]{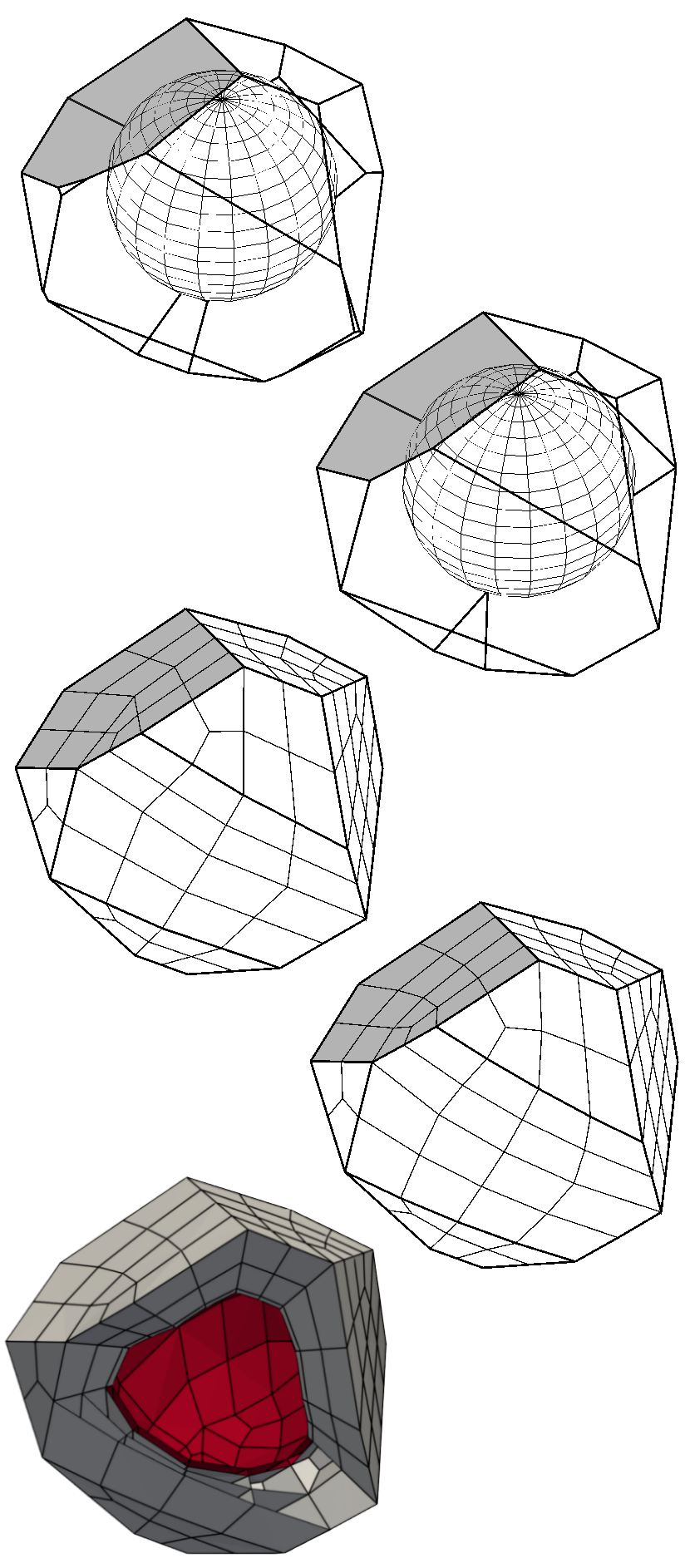}}
\end{picture}} \caption{A cell at different meshing stages. From top to bottom: 
Voronoi cell, edge-collapsed cell, all-quad surface, smoothed all-quad surface, 
and cut-away view showing swept hex elements.
\label{fig:cell} } \end{figure}

While the interstitial space in randomly packed spheres is complex,
several features of this problem make it possible to
recast the meshing question into a sequence of simpler, {\em local},
problems, making the overall problem far more tractable.  First, the
presence of so many surfaces provides a large number of termination points for
a given mesh topology.  Such surfaces occlude incident swept volumes, thus
bypassing the need to conformally merge advancing fronts and yielding
considerable savings in element count.  Second, decomposing the domain into
Voronoi cells defined by the sphere centers directly localizes the meshing
problem in several ways.

First, we reduce the problem to that of building a mesh that fills
the gap between the facets of the Voronoi cell and the sphere surface.
This process entails tessellating each Voronoi facet into an all-quad
decomposition and projecting these quads onto the sphere surface.  The
convexity of the Voronoi cell and coplanarity of the facets ensure
that this decomposition is possible.  Where desired, refinement in the
radial direction is always possible without disturbing other cells.

Second, tessellation of each Voronoi facet is a local problem.  Because
of bilateral symmetry about the Voronoi facet, tessellation of a facet
that is valid for one sphere will also be valid for the sphere on the
opposite side.   We note that facets may have an odd
number of vertices.  If, for example, the facet is a triangle, then
the all-quad tessellation will require midside node subdivision, resulting
in the introduction of new nodes along each edge.  To retain the
locality of the algorithm, we introduce midside nodes
on {\em each edge} of {\em each facet} throughout the domain.  With an even
number of vertices thus guaranteed, we can generate all-quad tessellations 
of the resulting polygons.

The strategy outlined above forms the essence of the proposed algorithm, and
several of the steps are illustrated in Fig. \ref{fig:cell}.  In
principle, it will produce a base mesh with relatively few elements that
inherit reasonable shape qualities from the Voronoi decomposition.  
In the following sections we dascribe
several important modifications to put the method into practice:
We mention these briefly as
edge collapse (to remove small facets);
vertex insertion on long edges (to balance the resolution);
facet tessellation and sweeping to the sphere surface;
mesh refinement; mesh smoothing; and
surface projection (to ensure that the final SEM nodal points are on the sphere
surfaces and boundaries while avoiding mesh entanglement).  
Save for the last step and (potentially) the smoothing step, all of the steps
of the algorithm are implemented in Matlab in $O(\cN)$ time, where $\cN$ is the
number of spheres.   Projection of the SEM nodal points and mesh smoothing are
performed in parallel by  using the open source spectral element code Nek5000.

Throughout, we assume the spheres are not touching.  To avoid a contact
singularity the spheres inflate towards their nominal touching radius,
we flatten the surfaces near contact points.  It is also relatively
simple to have the spheres in contact with a small fillet around the contact
point.  Such an approach is under development and illustrated in 
Fig. \ref{fig:contact}.

Elements of this work were inspired by 2D random media simulations
of Cruz, Patera, and co-workers.  Specifically, in \cite{cruz_patera_95}, these
authors introduce the idea of a disk-centered Voronoi decomposition for parallelism
and unstructured meshing.  In \cite{cruz_ghaddar_patera_95}, they discuss the
consequences of flattening or bridging touching disks to avoid the contact singularity.
Voronoi decompositions have been used in many other meshing applications as well.
Sheffer  et al.~\cite{sheffer_voronoi_99} describe application of embedded
Voronoi graphs to decompose geometries into sweepable domains.
Yan  et al.~\cite{levy2010} develop efficient algorithms for constructing
clipped Voronoi diagrams and applying these to tetrahedral mesh generation.

%
%
%
%

We organize the paper as follows.
In Section 2 we present algorithmic details.
In Section 3 we demonstrate various pebble meshes, pebble-bed reactor
simulations, and performance on Summit.
We conclude with remarks and discussion in Section 4.

\section{Algorithm}      
Our principal objective is to produce all-hex meshes with
roughly 300--400 elements per sphere so that the required 
simulation resolution can be realized through efficient high-order 
(e.g., $N=7$ to 9) polynomial approximations that are the foundation of
the SEM.  The SEM basis consists
of $N$th-order Lagrange polynomials based on tensor products of the 
Gauss--Lobatto--Legendre (GLL) quadrature points in the reference element,
$\Oh:=[-1,1]$, which is isoparametrically mapped to $E$ curvilinear hexahedral
(hex) elements.  The effective resolution is  $n=EN^3$. With a target value 
of $n$ required to resolve the important turbulent length scales, one can
either increase $E$ or increase $N$.  Through decades of experience, we
have found that $N=7$ is a nearly optimal value for Nek5000 because it
realizes high throughput with reasonable element counts and reasonable timestep
sizes~\cite{fischer20a}.

Throughout this article, we will consider the specific case of a computational
domain $\Omega$ that is a cylinder of fixed height $z=H$ and radius $R_c$, minus
the space occupied by $\cN$ spheres of unit radius, $R$.  Other configurations
are of course possible, but this geometry will suffice to describe the
basic approach.  

\subsection{Voronoi Diagram}

The starting point for our algorithm is a user-provided set of sphere centers,
$\cP := \{ \bp_i \}$, $i=1,\dots,\cN$, which are typically obtained from
experiments or from a discrete element method (DEM).  The user may or may
not also provide domain boundary information, which needs to be verified in any
case in order to provide a precise mesh (i.e., one in which the spheres, with
their nominal radii, actually touch the boundary to within the prescribed
tolerance).\footnote{We initially identify the cylinder radius, $R_c$, and 
center, $\bx_c=(x_c,y_c)$, by solving for the center position that minimizes the
$p$-norm, $\| {\bf \up}-\bx_c \|_p$, over the set of sphere centers ${\bf \up}$.
A large value of $p$ (e.g., $p$=100) approximates the infinity norm.  Two
passes are made---one with all sphere centers, and then one with spheres that
are within $2R$ the estimated cylinder boundary.}

\begin{figure}[t] \centering {\setlength{\unitlength}{1.0in}
\begin{picture}(2.300,2.50)(0,0)
\put(-.05,0.03){\includegraphics[width=2.3in]{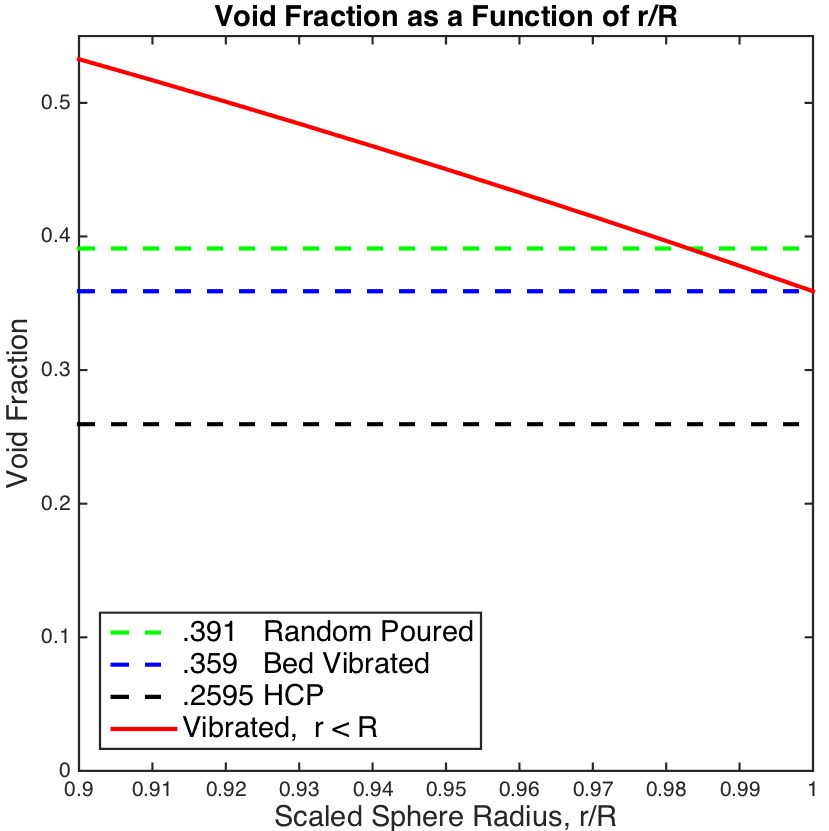}}
\end{picture}} \caption{Void-fraction as a function of sphere radius.
\label{fig:voidf} } \end{figure}

The central element of our meshing scheme is the set of Voronoi cells
that bound each sphere. For spheres that are near the domain boundary, the
Voronoi cells will extend to infinity unless they are clipped.  While
clipped Voronoi algorithms are well established (e.g., \cite{levy2010}),
we are using the Voronoi utility in Matlab as a black box.  To trim
the Voronoi diagram, we augment $\cP$ with additional
sphere centers outside $\Omega$, which we call {\em ghost spheres}.  To
generate this auxiliary set in the case of a cylindrical domain, we first
reflect any point within a distance $\Delta_r < 2R$ of the cylinder
wall to a new position that is at radius $R_c + \Delta_r$ along the same radial
coordinate.  From this augmented set, we take each center point that is at a
height $z < 2R$ and reflect it about $z=0$.  Similarly, we take all points at
heights $z > H - 2R$ and reflect these about $z=H$.   From this augmented
point set, ${\bar \cP}$, we generate the Voronoi cells by calling Matlab's
{\tt voronoin} function with the {\tt 'C-0'} argument to remove redundantly
represented vertices.  We subsequently restrict our interest to the first $\cN$
cells.  We note that the runtime for the Voronoi decomposition is $O(\cN)$
except in pathological cases (e.g., a crystalline lattice).

\subsection{Preconditioning the Data}

\begin{figure}[t] \centering {\setlength{\unitlength}{1.0in}
\begin{picture}(2.300,5.20)(0,0)
\put(-.05,2.55){\includegraphics[width=2.63in]{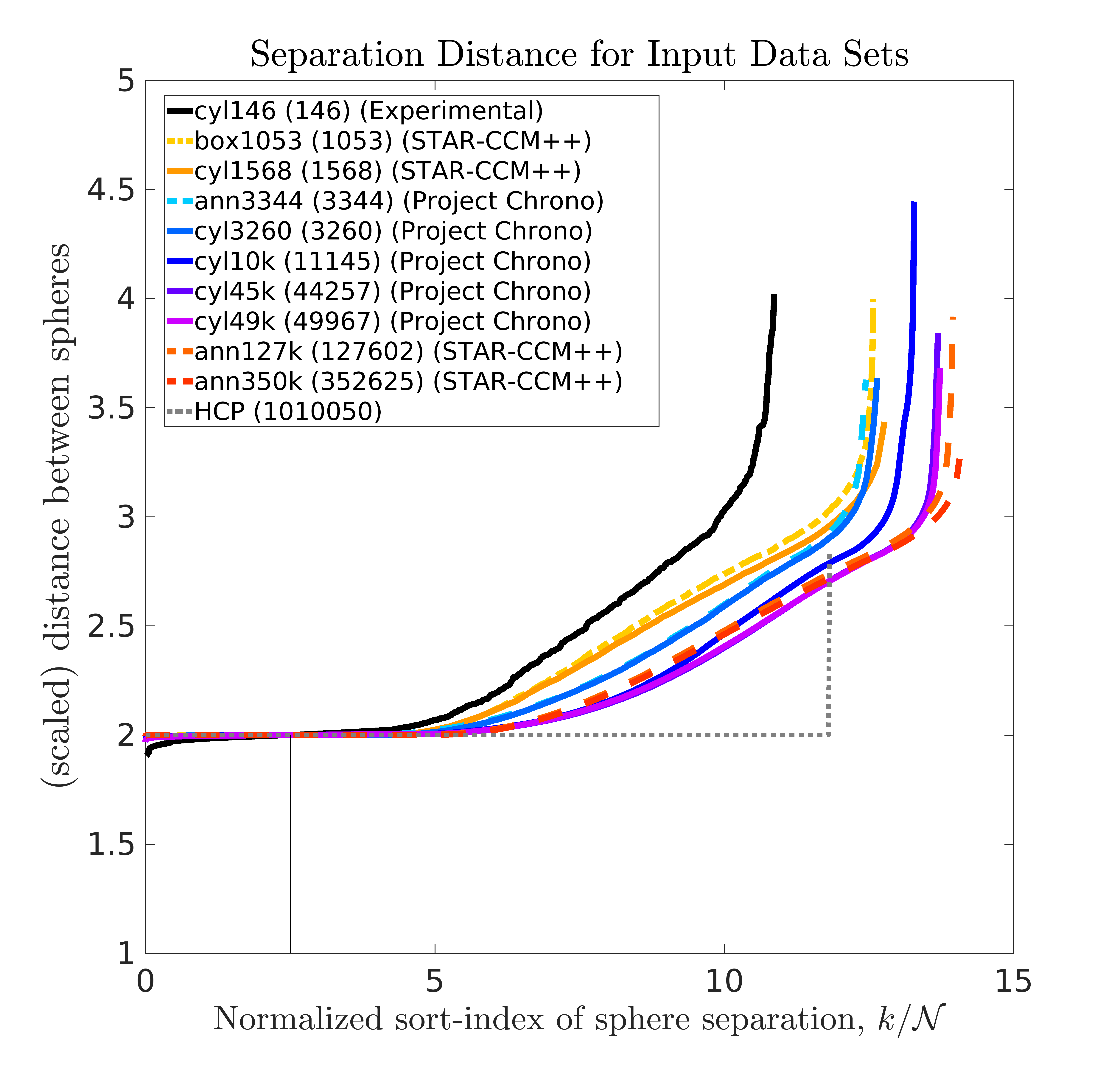}}
\put(-.05,-.05){\includegraphics[width=2.63in]{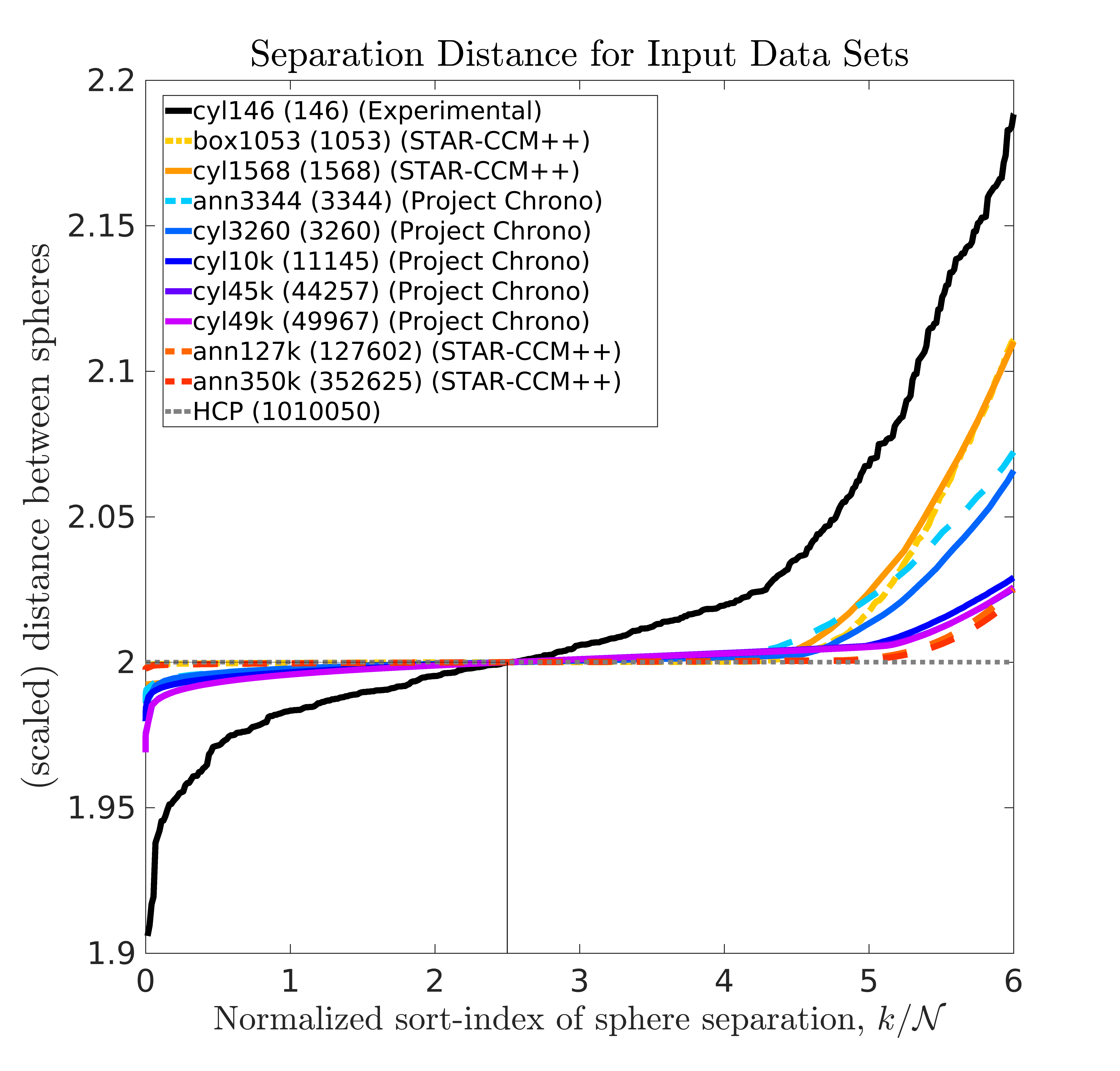}}
\end{picture}} \caption{ Top: sorted list of neighbor-neighbor distances for
$\cN=146$ to 350K.  Bottom: zoom near $\Delta_{ij}=2.$ 
\label{fig:distance} } \end{figure}

One challenge of meshing packed beds is the prevalence of point-contact
singularities where the spheres touch.  From a flow perspective, the
fluid motion is almost nil in these regions so having the precise geometry
near the contact is not requisite.  One can avoid the singularity
by slightly reducing the radius, introducing a flat-spot near the
contact point, or introducing a solid {\em bridge} that connects
the spheres ({\rm e.g.}, Fig. \ref{fig:contact}).
The latter approaches were considered in 2D in \cite{cruz_ghaddar_patera_95}.
For the shrinking case, we consider spheres of radius $r < R$, 
where $R$ is one-half the nominal separation of the sphere centers.
The overall pressure drop is strongly dependent on void fraction, which
in turn is strongly dependent on $r/R$, as illustrated in Fig.
\ref{fig:voidf}.  We see that at $r/R=0.95$, the void fraction 
is 25\% too large compared with the $V_f=0.359$ value for a vibrated poured bed of
spheres.  Also shown are the values for a random-poured bed, $V_f=0.391$, and
the hexagonally close-packed (HCP) case, which attains the minimum possible
value of $V_f=0.2595$.

Given the sensitivity of the pressure drop to void fraction and ultimately
to the sphere radii, it is important to accurately identify the nominal
   separation of the spheres from the given data set, $\cP$.
Let $\Delta_{ij} := \|\bp_i - \bp_j\|$ be the Euclidean separation for all
$i$-$j$ pairs connected by the Delaunay triangulation of $\cP$.  
If the data set were perfect, the target radius would simply be $R = R_{\min}
:= \frac{1}{2}\min \Delta_{ij}$.  However, virtually all data sets have a
distribution in which a handful of element pairs are closer than others, which
implies that choosing $R=R_{\min}$ would yield too large of a separation almost
everywhere. 

We identify a more robust separation definition as follows.
Sort the list $\{ \Delta_{ij} \}$ in ascending order and plot
these values, as shown in Fig. \ref{fig:distance}.
We expect there to be a plateau in this sorted list corresponding to
the pairs of spheres that are touching.
The cardinality of $\{ \Delta_{ij} \}$ is $O(\cN)$, so it makes
sense to scale the $x$-axis by $1/\cN$ so that graphs for different data
sets can be plotted in the same figure.\footnote{Note that for the HCP case
we expect to have $\leq$ 12 connections per sphere (i.e.,
for each $i$, there will be $\leq$ 12 nontrivial entries, $\Delta_{ij}$),
with inequality resulting from spheres at the domain boundary that have
fewer than 12 connections to other spheres.}
With this scaling, we see that the ten data sets in Fig. \ref{fig:distance}
exhibit plateaus on the interval $k/\cN \approx [0,5]$. The cyl146 case,
corresponding to measured experimental data \cite{yassin}, has the
least well-defined plateau.  From these collective sets we choose the value of
$\Delta_{ij}$ corresponding to ranking $k = 2.5 \cN$ as the nominal separation,
which we denote as $\Delta^*$.  The first step in our algorithm is to rescale
the input geometry by $\frac{1}{2}\Delta^*$ so that the target radius is $R=1$.
This scaling has been applied in Fig. \ref{fig:distance}.

\begin{figure*}[t] 
\centering
  \subfigure[Group edges]{
    \includegraphics[trim=150 0 150 0,clip,width=0.13\textwidth]{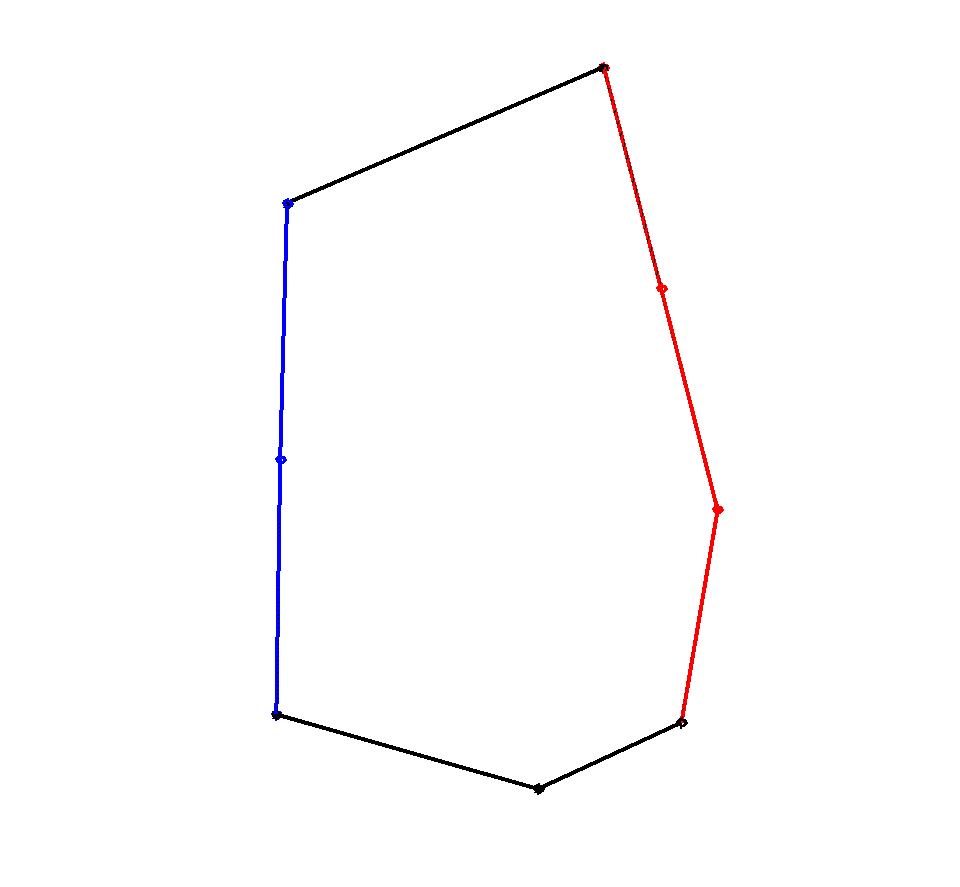}
  }
  \subfigure[Find sections]{
    \includegraphics[trim=150 0 150 0,clip,width=0.13\textwidth]{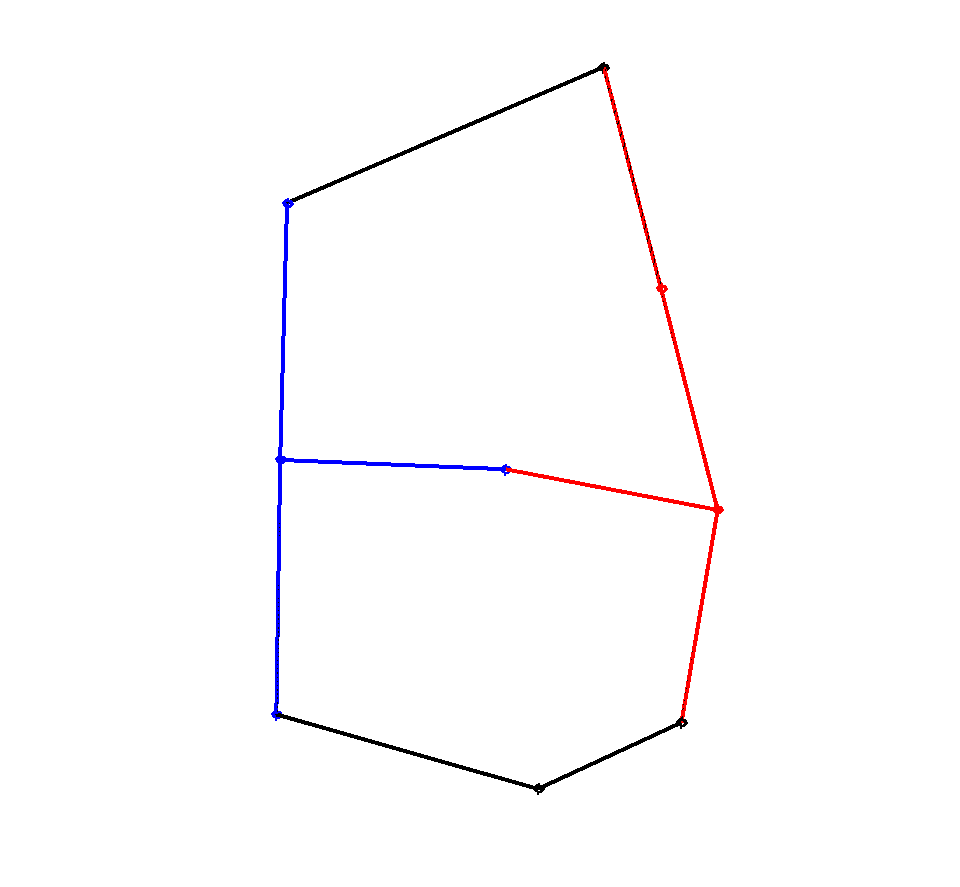}
  }
  \subfigure[Divide and conquer]{
    \includegraphics[trim=150 0 150 0,clip,width=0.13\textwidth]{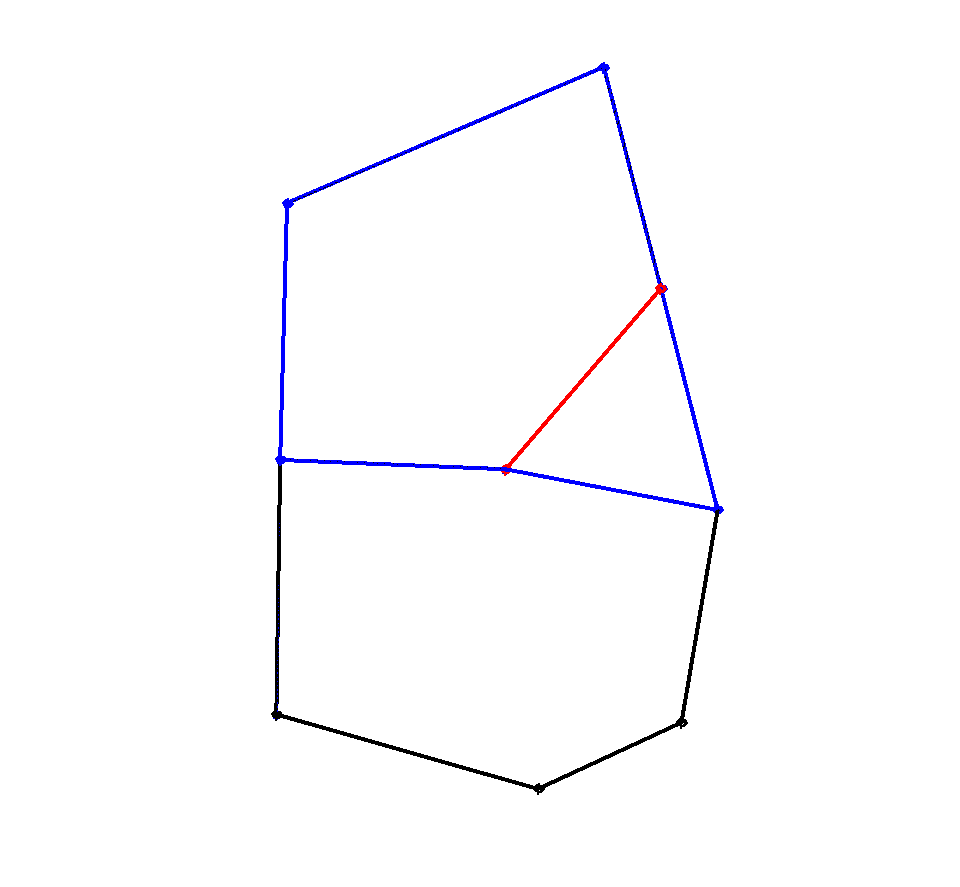}
    \includegraphics[trim=150 0 150 0,clip,width=0.13\textwidth]{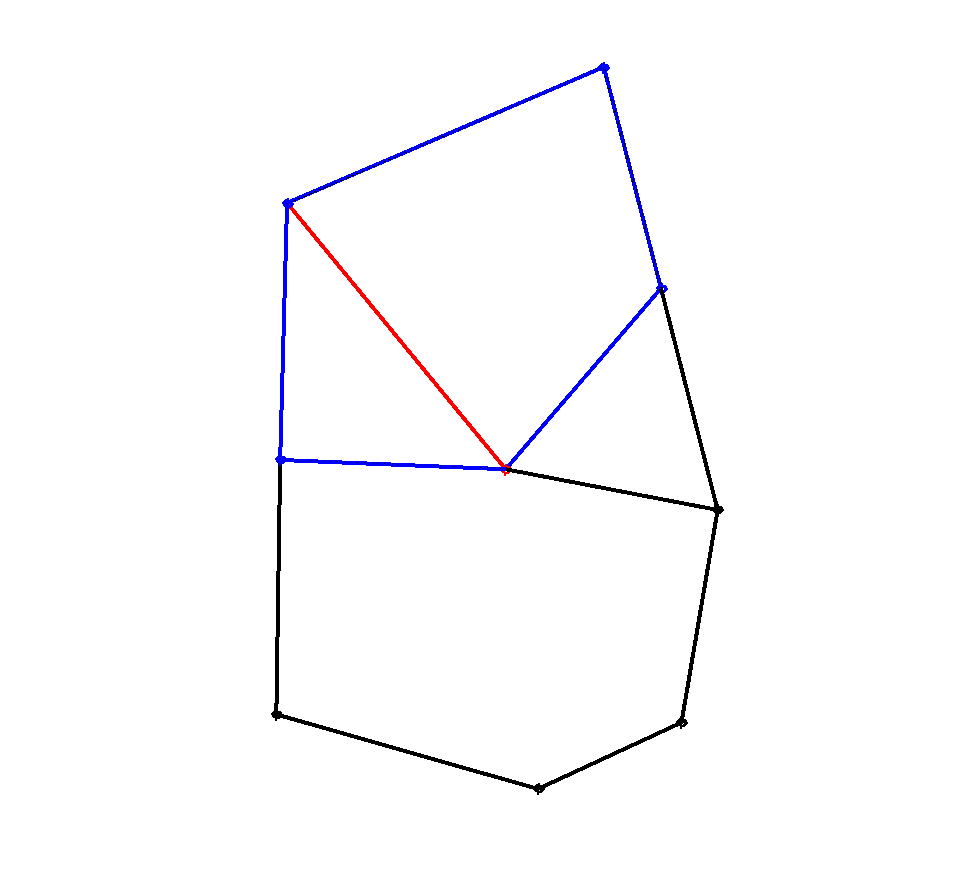}
    \includegraphics[trim=150 0 150 0,clip,width=0.13\textwidth]{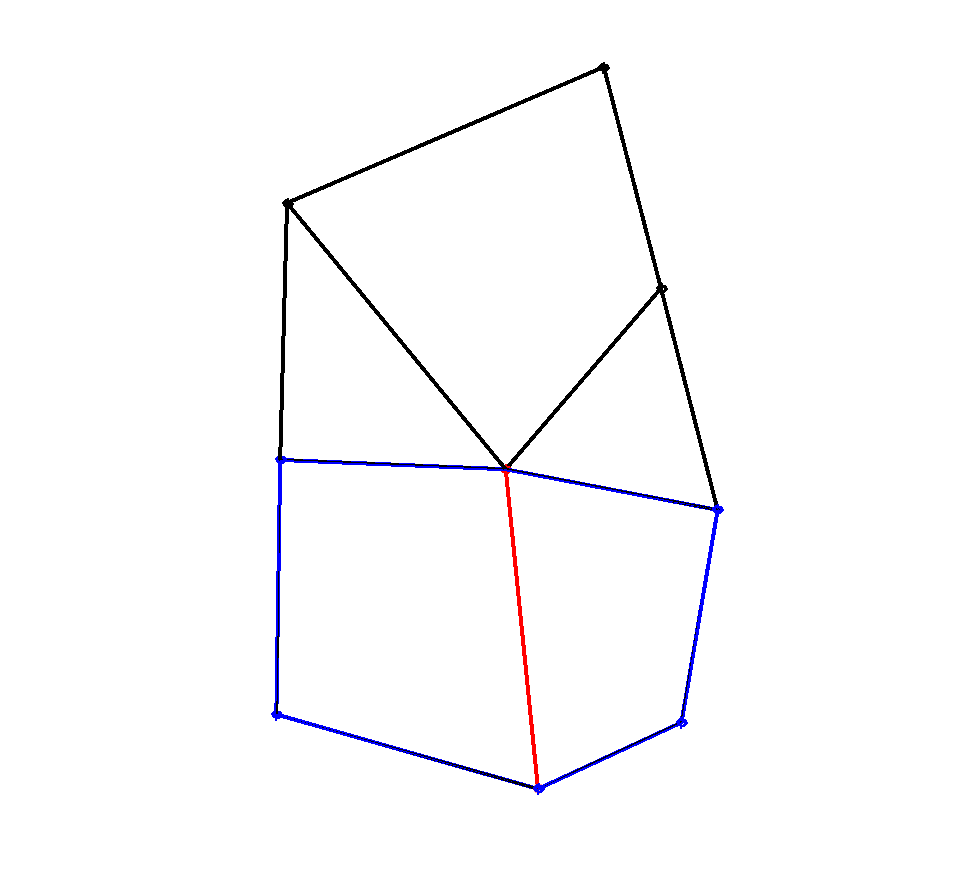}
  }
  \subfigure[Quads unsmoothed/smoothed]{
    \includegraphics[trim=150 0 150 0,clip,width=0.13\textwidth]{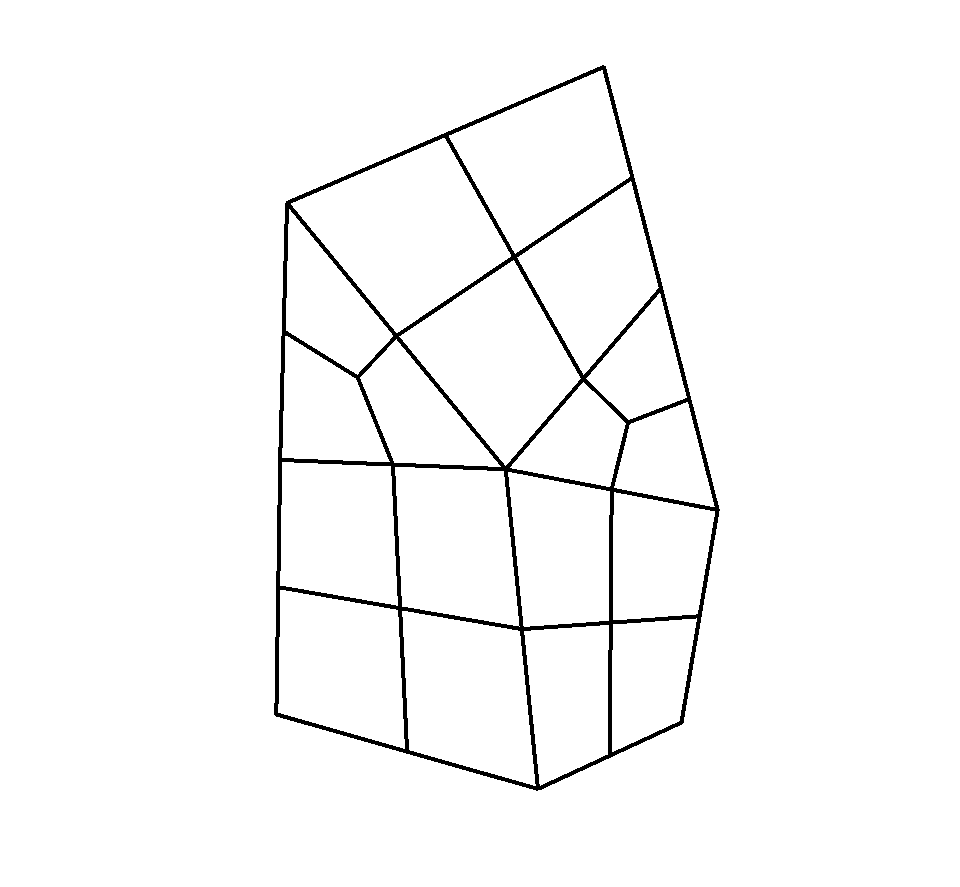}
    \includegraphics[trim=150 0 150 0,clip,width=0.13\textwidth]{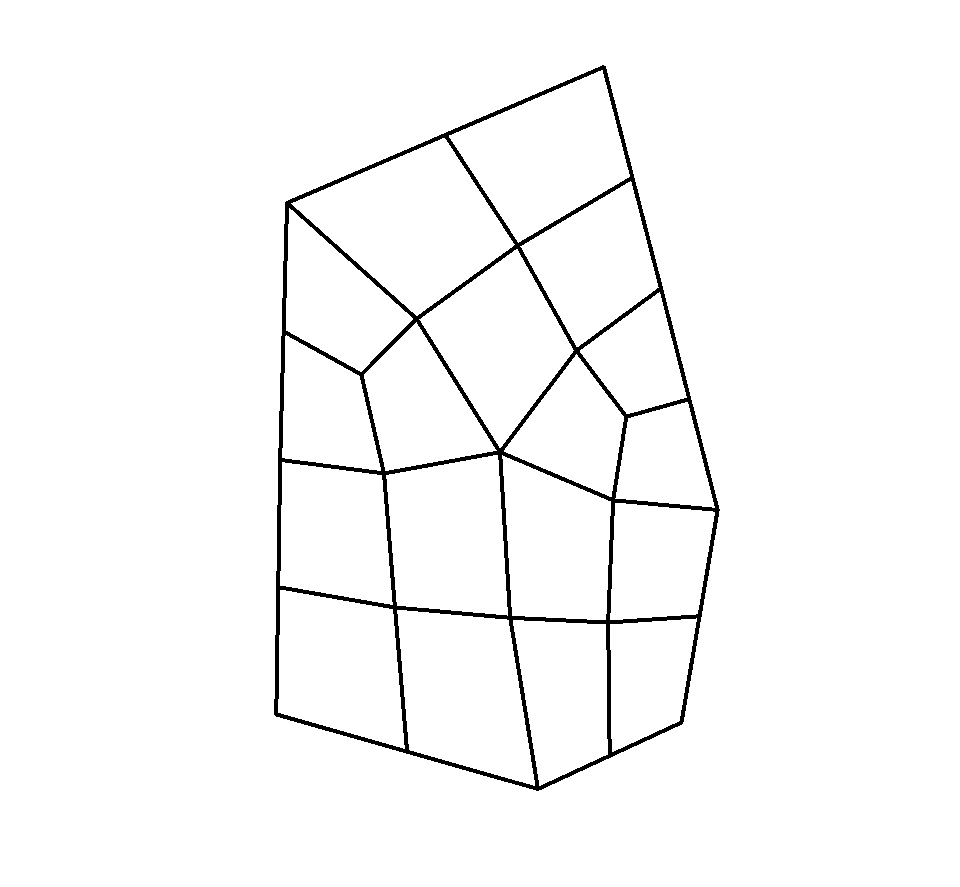}
  }
\caption{Facet tessellation steps.
\label{fig:quad} }
\end{figure*}

\subsection{Edge Collapse}
Despite its provably excellent properties of convexity with convex planar
facets, the Voronoi tessellation may still contain very thin facets (slivers) 
that can lead to poorly conditioned elements.  To eliminate these and
to generate a more uniform mesh with a bounded ratio of the longest to the
shortest edge, we first perform edge collapse on the Voronoi cells.  Any edge
shorter than a given tolerance is collapsed, and its two vertices are 
fused into one.  If after edge collapse the number of edges on a facet is $<$ 3, 
the facet is deleted.   Subsequent to edge collapse, we use {\em vertex insertion}
to ensure that the {\em longest} edge is below a certain threshold.

A risk with edge collapse is that facets lose their planarity and, worse,
might not face the sphere that they are nominally bounding.  We take several
steps to avoid this scenario.  First, our target edge collapse tolerance
of $tol_{\infty}=0.35R$ is not overly aggressive.   Second, we make several
passes through the data, collapsing the shortest edges first and not allowing
a single vertex to be moved more than once per pass, even if it is attached
to more than one short edge.  With each pass, we set the tolerance to be
$tol_k = 0.35R (0.6)^{8-k}$ for $k=1,\dots,8$, and $tol_k=0.35R$ for $k$=9 and 10.
As a final precaution, we have the option of revisiting a neighborhood if at the
end of the meshing process the algorithm produces hexes with inverted Jacobians.
In that case we rerun the algorithm with tighter local tolerances.  In the limit
of $tol_{\infty}=0$ we recover the provably workable properties of the original
Voronoi tessellation.   (We find this corrective step necessary only for the
$\cN=350$K case.) We remark that many of the challenges in the meshing
process come from the domain boundaries, where the Delaunay triangulation has
relatively long edges.   Consequently we typically set $tol_{\infty}=0.25$ in
the neighborhood of $\dO$.

Another issue with edge collapse is that the reconstructed facet surface will
potentially intersect the sphere that is to be bounded by the facet.  To
avoid this situation, we project all points generated during edge insertion
or facet tessellation onto a sphere that is larger than $R_0$. 

\subsection{Vertex Insertion} 
In addition to having short edges (mostly cleaned by edge collapse), the
initial Voronoi decomposition can yield facets with edges that are longer
than desired.  For this reason we insert vertices for edges that are
longer than $0.8R$; thus we have
a maximum edge-length ratio of 0.8/0.35.  This value is readily adjusted
at construction time and is also adjusted somewhat by the final mesh smoothing
process.  For our initial trials, however, it seems to be a reasonable
ratio.

\subsection{Facet Tessellation}

A major part of the algorithm is the all-quad decomposition of the Voronoi facets.
Each facet can be tessellated into a set of quadrilaterals by first inserting a
vertex at the midpoint of each edge in order to guarantee that the facet is a
polygon with an even number of edges.  Alternatively, one can decompose the
facet into quads and triangles and perform the midside node insertion at the
end.  We follow the latter approach.

As illustrated in Fig. \ref{fig:quad}, our facet tessellation uses a two-phase
divide-and-conquer algorithm to reduce the number of vertices on each polygon
by splitting facets into smaller polygons.  Phase one, illustrated in 
Fig. \ref{fig:quad}(a) and (b), begins by clustering sequences of edge vertices
    having angles $> 155^{\circ}$ into {\em edge groups}.   
If the facet is large (having either  many vertices or large area), we 
insert a point at the barycenter.  Each edge group having at least one
interior vertex angle will connect one and only one vertex to the
barycenter.  The selected vertex is the one with
minimal bias between its adjacent angles (i.e., as close to a bisector as
possible).  In the case of Fig. \ref{fig:quad}(b)
we see that this process subdivides the facet into two subdomains.

For each subdomain (or for the whole facet if we skip phase one), 
we consider sequences of 4 successive vertices to see whether they produce
viable quads and whether the remaining part of the domain meets quality
conditions such as low variance in edge length and absence of large
angles.   We also consider finding triangles from successive vertex
triplets with similar quality metrics.  With a slight bias toward 
quads, the quality indices are compared, and the algorithm chooses the
partition that yields the best quality.  
In the upper subdomain of Fig. \ref{fig:quad}(c) 
we see that a triangle is selected that is nearly equilateral.
The algorithm is applied recursively, which yields in this case
another triangle and a quad in the upper subdomain.   Applying the
same algorithm to the lower subdomain yields two quads.

From the divide-and-conquer phase, all quads are subdivided into four 
smaller quads and triangles into three quads, resulting in an
initial all-quad tessellation of the facet, shown in
Fig. \ref{fig:quad}(d), left.
Laplacian smoothing is applied to this decomposition
to yield the final result, shown in Fig. \ref{fig:quad}(d), right.

We reiterate that, because of edge collapse, the edge points are typically not
planar, and points in the facet interior are therefore projected onto the
original bisecting plan to avoid ending up in the sphere interiors.  The
full-mesh smoothing step will accommodate any misshapen elements generated by
this projection.


\subsection{Sweeping}


Once the facets are tessellated, we generate an initial all-hex mesh
by projecting each facet-based quadrilateral onto the sphere(s) that
are bounded by the facet, or onto the domain boundary.  To 
leave room for an additional thin element layer (``boundary layer'')
near the spheres, we take the initial sphere diameter to be $R_0 = 0.8889R$.
The sweeping is illustrated in Fig. \ref{fig:cell}.
In this initial phase, however, only one layer of elements is generated 
so that an initial mesh smoothing may be applied at relatively low element
counts.


\subsection{Mesh Smoothing}
\label{sec:smooth}

The initial all-hex mesh (comprising 33M elements in the $\cN$=350K
case), is smoothed by using a combination of Laplacian smoothing and
element-Jacobian optimization following the strategy outlined in
~\cite{mittal19a,knupp1}.  In addition to the condition-number-based
optimization function, we add a substantial penalty term for negative
Jacobians.  The mesh smoothing thus consists of two parts: Laplacian smoothing
and optimization.  These are alternated over a sequence of ten passes.  Because
of the complexity of the objective function, we execute this phase in parallel,
using Nek5000 \cite{mittal19a}.  We describe the basic smoothing steps in the
sequel.

\subsubsection{Laplacian smoothing}

Nek5000 has a highly scalable gather-scatter utility, {\em gslib}, which is
a stand-alone C library that scales to millions of ranks.  It is the
workhorse for all interelement operations in Nek5000 because it requires no
topological information other than a global ID for each vertex in the graph.
(There are 8 such vertices for each hex, each having
a unique ID in the global mesh.)   The Laplacian smoother is built on
{\em gslib}'s vector assembly operation, which effects a sum and redistribute
of values sharing the same global IDs in the graph.
The mathematical expression for this operation is $\uv_L = QQ^T {\tilde \uv}_L$,
where $Q$ is a Boolean matrix that maps (copies) global entities to their
local (element-based) counterparts (the FEM scatter operation) \cite{dfm02}.

We begin with a shrinking step.
For each element, $\Omega^e$, we have an isoparametric mapping of the form
\begin{eqnarray}
\bx \bigg|_{\Omega^e} \;=\;
   \bx^e(\br) &=& \sum_{i=1}^8 \bx_i^e l_i(\br),
\end{eqnarray}
where $\br \in \Oh := [-1,1]^3$ and $l_i$ is the set of cardinal Lagrange
basis functions having nodes at the 8 vertices, $\br_i$, of $\Oh$.
For each element we create a new set of coordinates,
\vspace{-.05in}
\begin{eqnarray}
   {\tilde \bx}^e_j &:=& \sum_{i=1}^8 \bx_i^e l_i({\tilde \br_j}),
\end{eqnarray}
where ${\tilde \br_j} = s \br_j$ and $s=0.95$ is a scale factor.
The new coordinates are thus interpolants of $\bx^e_i$, {\em interior 
to $\Omega^e$}.  On domain boundaries, we set the corresponding
${\tilde \br}=(\tilde{r},\tilde{s},\tilde{t})$ to $\pm 1$ as needed, so that
the points do not move normal to the domain surface.

Subsequent to this shrinking step, we apply direct stiffness summation
\cite{stfi73} (actually, direct stiffness averaging), in which the
coordinates of elementwise-shared vertices are added together and 
divided by the vertex multiplicity (i.e., the number of elements that
share that global vertex).  We then reproject boundary points to their
corresponding boundary surface since the averaging step may potentially 
move vertices off of curved surfaces.  This Laplacian smoothing
process is fast, both in serial and in parallel.  It can be repeated
multiple times for varying values of $s$.  We typically apply it about ten
times in order to even out the element sizes, particularly on the sphere surfaces.

\subsubsection{Optimization smoothing}

The second smoothing tool is based on optimization, following the ideas 
of Knupp \cite{knupp1,knupp2,knupp3} and Mittal and Fischer \cite{mittal19a}.
Unlike Laplacian smoothing, it is more localized and will not tend to 
redistribute the resolution.  It will, however, ensure locally high-quality
elements.

We define as an objective function the condition number of the local 
Jacobian matrix in the Frobenius norm. For each vertex, $\bx_i^e$,
the local Jacobian matrix is 
\begin{eqnarray}
\left[J^e_i\right]_{j,k} = \pp{x_j}{r_k}\bigg|_{\bx=\bx^e_i},
\end{eqnarray}
where $\bx=(x_1,x_2,x_3)$ is the physical coordinate and 
$\br=(r_1,r_2,r_3)$ is the reference coordinate.
The global objective function is the average of local functions,
\begin{eqnarray}
f(\ubx) = \frac{1}{E} \sum_e^E {\tilde \phi^e} 
= \frac{1}{E} \sum_e^E\left[\frac{1}{8}\sum_i^{8}(\phi_i^e)^2\right], 
\end{eqnarray}
with
\begin{eqnarray}
\phi_i^e = \frac{1}{3}\left(||J_{i,e}||_F||J_{i,e}^{-1}||_F\right).
\end{eqnarray}
The minimum value of the global objective function is 1, which
is realized when the mesh is a 3D cubic lattice. 

To repair elements that have a negative Jacobian, 
we augment ${\tilde \phi}^e$ with a penalty term,
\begin{eqnarray}
\hspace{-.05in}
\hat{\phi^e}(\bx^e) &=&
{\tilde\phi^e}(\bx^e) - 
\tau \, J_i^e/J_{max}\cdot\bbone_{\{J_i^e<\epsilon\}}(\bx^e).
\hspace{.05in}
\end{eqnarray}
We choose $\tau=1000E$, $\epsilon=0.001$, and $J_{max}=\max_{i,e} J_i^e$.

The optimization is solved by the conjugate gradient method. 
For each global vertex $i$, the $j$th dimension gradient of the 
objective function can be computed by 
\begin{eqnarray}
g_j(\bx_i) 
&=&
\sum_{e=1}^E \pp{\hat{\phi^e}}{x_j}\bigg|_{\bx_i} 
\;=\;
\sum_{e\in\cE(x_j)}\pp{\hat{\phi^e}}{x_j}\bigg|_{\bx_i},
\end{eqnarray}
where $\cE(\bx_i)$ is a set of elements sharing the vertex $\bx_i$.
The last sum can be assembled by $Q^T$, the gather (i.e., direct-stiffness
summation) operation in the FEM, such that $\ug = Q^T \ug_L$. This framework
allows us to compute the gradient locally, element-by-element, and to then
gather all local gradients at once.

We approximate the local gradient with a central difference approximation
along the unit vector $\be_j$ in the $j$th dimension,
\begin{eqnarray}
g^e_{i,j} &\approx& 
\frac{1}{2h}\left[\hat{\phi^e}(\bx_i+h\be_j)-\hat{\phi^e}(\bx_i-h\be_j)\right],
\end{eqnarray}
where the step size $h$ is chosen to be $0.001$ times the shortest edge.

We remark that for large element counts (e.g., $E=10^5$--$10^8$),
our Matlab version of optimization is slow because
it is difficult to avoid {\em for} loops.  We thus execute the mesh smoothing
in Nek5000 since these routines are already available there \cite{mittal19a}.  
Since the quality of the initial mesh coming from the Voronoi-based scheme is
already high, we do not have difficulty with mesh tangling.  Nonetheless, this
is one area where our algorithm could be improved by using, for example,
edge-oriented algorithms such as presented in \cite{edge1,edge2}.

\subsubsection{Boundary smoothing}

The mesh smoothing or optimization is constrained by the boundary condition. 
For improved mesh quality, we allow the boundary vertices to slide 
along the boundary surfaces. As mentioned earlier, for Laplacian smoothing, the
direction toward the boundary face will not be shrunk, and points are reprojected onto
the surface after the averaging operator.

As for the optimizer, the full objective function should include a Lagrange 
multiplier that would add a penalty for boundary points moving away from boundary. 
In fact, the gradient of the boundary constraints are in the same direction of the 
normal vector of the boundary surface. Therefore, we decompose the gradient
at the boundary into normal and tangential components and  have to eliminate 
the movement only along the normal direction,
\begin{eqnarray}
\hat{\ug}_{e,i}\big|_{\dO} 
&=&
\ug_{e,i}\big|_{\dO} - \ug_{e,i}\big|_{\dO}\cdot{\bhn}.
\end{eqnarray}
To make sure the boundary points are still attaching on the boundary surface after 
all of the movement, we apply the orthogonal projection at each iteration.

\subsection{Mesh refinement}
\label{sec:refine} 

The mesh smoothing will produce a valid mesh. At this stage, in each cell, 
there is only one layer of elements between facet and sphere. In order to meet the 
requirement for the fluid simulations, the mesh needs to be refined.

\subsubsection{Cell refinement}
In each cell, we can arbitrarily refine the element in the radial direction
of the sphere without breaking the conformality between other cells. Here,
we add a midlayer by splitting the elements into two in the radial 
direction. To gradually increase the resolution near spheres, we split the 
elements in a ratio of 55\% (near facet) versus 45\% (near sphere).
This action effectively doubles the number of elements.

\subsubsection{Extrusion}
Extrusion is used to generate extra layers of elements for CFD computations. 
We typically generate one extra layer inward to the spheres to resolve the 
boundary-layer of the fluid solution near the sphere walls. 
Similarly, we add an extra layer outward on the cylinder wall.
In the flow direction we extrude three layers toward the bottom of the domain
(the flow inlet) and seven layers for the outflow region on the top.
The single-layer extrusion on the spheres increases the total number of
elements by an additional $1.5 \times$ after the first refinement.
As seen in Fig. \ref{fig:cell}, the resulting mesh has three layers 
between the facet and sphere.  
For polynomial degree $N=7$, this yields about 42 points between 
adjacent spheres, which is sufficient for large-eddy simulation at
the target Reynolds numbers based on the hydraulic diameter.\footnote{The
Reynolds number $Re=UD_h/\nu$ is the flow velocity, $U$, nondimensionalized
by the hydraulic diameter of the passageway, $D_h$, and kinematic viscosity,
$\nu$ of the fluid.}

\subsection{Projection}
\label{sec:project}

\begin{figure}[t] \centering {\setlength{\unitlength}{1.0in}
\begin{picture}(2.300,2.00)(0,0.0)
\put(-.00,0.00){\includegraphics[width=2.30in]{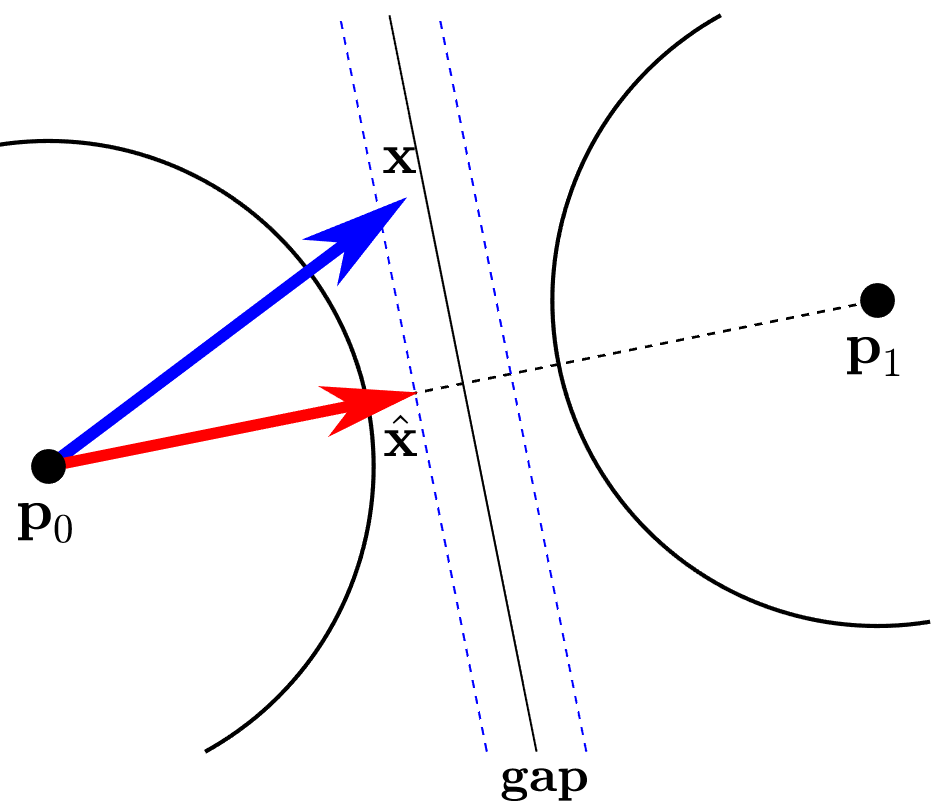}}
\end{picture}} 
\caption{Enforcement of a gap between Sphere 0
and 1.  Any point on the surface of Sphere 0 that is nominally
projected to $\bp$ is shortened so that its projected component onto
$\bx_0$--$\bx_1$ is equal to $\hat \bp$.
\label{fig:gap} } \end{figure}

The final step of the meshing process is to project the facet quadrilaterals
that tessellate the sphere and boundary surfaces onto their actual locations.
This step is done inside Nek5000 because it requires projection of the full
set of GLL points onto the target geometry.  Typically, we perform this step
in two passes.  First, we project at low-order ($N=2$) in order to generate
a hex27 mesh description that can be used as a starting point for Nek5000.
In this phase, we also inflate the sphere surfaces to the target radius, $r < R$. 
Flat spots are included to enforce a prescribed gap that avoids contact
singularities using the projection algorithm illustrated in Fig.  \ref{fig:gap}.

\begin{figure*}
 \centering
 \includegraphics[width=0.95\textwidth]{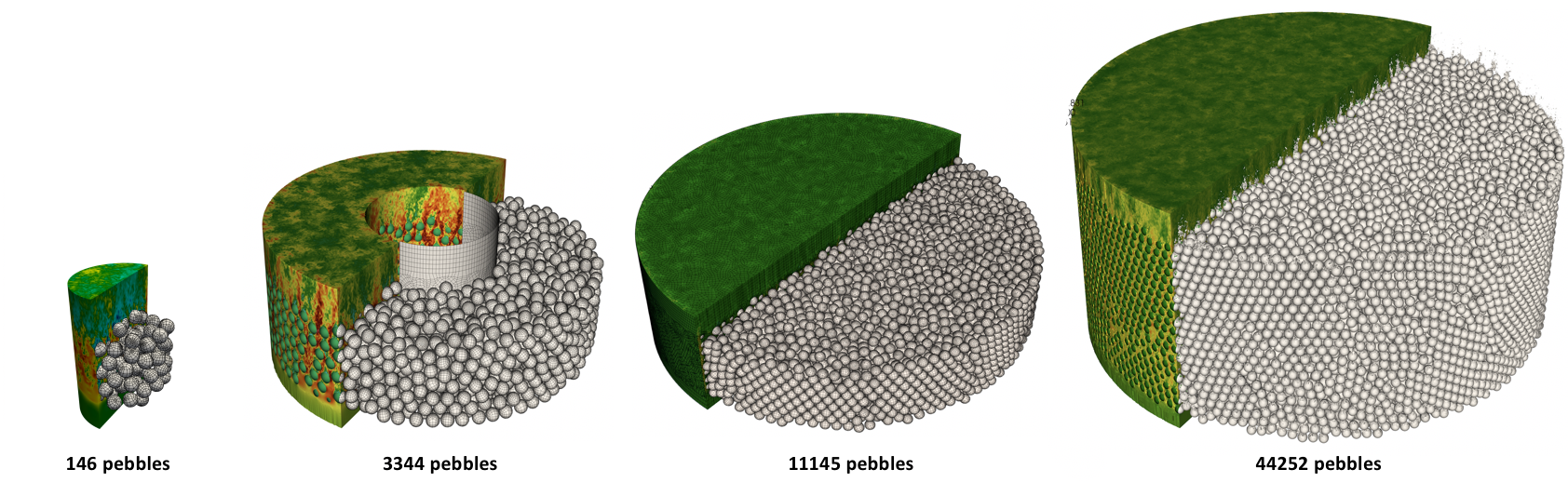}
 \caption{\label{fig:all_pebbles}
  Pebble meshes and simulations.}
\end{figure*}

\begin{table*}
\begin{center}
\footnotesize
\begin{tabular}{ |l|r|c|c|r|r| }
  \hline
  \multicolumn{6}{|c|}{{\bf Statistics for Several Multisphere Configurations}} \\
  \hline
  \hline
   Case    &\multicolumn{1}{c|}{$\cN$}
                     & Source         & Container &\multicolumn{1}{c|}{$E$}
                                                               & $E/\cN$ \\
 \hline
   cyl146  &     146 & Experiment     & cylinder  &     62,132 & 425.56 \\
   box1053 &   1,053 & StarCCM+       & box       &    376,828 & 250.72 \\
   cyl1568 &   1,568 & StarCCM+       & cylinder  &    524,386 & 334.43 \\
   cyl3260 &   3,260 & StarCCM+       & cylinder  &  1,121,214 & 343.93 \\
   ann3344 &   3,344 & Project Chrono & annulus   &  1,133,446 & 338.95 \\
   cyl11k  &  11,145 & Project Chrono & cylinder  &  3,575,076 & 320.78 \\
   cyl44k  &  44,257 & Project Chrono & cylinder  & 13,032,440 & 294.47 \\
   cyl49k  &  49,967 & Project Chrono & cylinder  & 14,864,766 & 297.49 \\
   ann127k & 127,602 & StarCCM+       & annulus   & 39,249,190 & 307.52 \\
   ann350k & 352,625 & StarCCM+       & annulus   & 98,782,067 & 280.13 \\
 \hline
\end{tabular}
\end{center}
\caption{\label{tb:cases} 
List of cases meshed to date including the number of spheres $\cN$,
the data source, domain shape, number of elements $E$, and number
of elements per sphere.}
\end{table*}

\begin{table*}
\footnotesize
\begin{center}
\begin{tabular}{|l|r|r|r|r|r|r|r|}
\hline
  \multicolumn{8}{|c|}{{\bf Meshing Time Breakdown (sec)}} \\
  \hline
Meshing Step & cyl146 & cyl1568 & ann3344 & cyl11k & cyl49k & ann127k & ann350k \\
\hline
\hline
  IO for Qhull
        & 4.56E-01 & 1.10E+00 & 2.64E+00 & 6.36E+00 & 1.97E+01 & 5.13E+01 & 2.79E+02\\
  Voronoi cells (Qhull)
        & 1.70E-01 & 4.29E-01 & 1.07E+00 & 2.50E+00 & 8.77E+00 & 2.12E+01 & 7.98E+01\\ 
\hline
\hline
  Facet generation
        & 1.66E-01 & 1.51R+00 & 4.22E+00 & 1.90E+01 & 1.77E+02 & 9.14E+02 & 4.20E+03\\ 
  Edge collapse
        & 8.67E-02 & 2.34E-01 & 4.53E-01 & 1.26E+00 & 5.24E+00 & 1.30E+01 & 8.20E+01\\
  Facet/edge clean-up 
        & 1.21E+00 & 6.57E+00 & 8.66E+00 & 2.75E+01 & 1.29E+02 & 3.47E+02 & 2.55E+03\\
\hline
\hline
  Tessellation 
        & 1.43E+00 & 9.76E+00 & 1.87E+01 & 6.26E+01 & 2.84E+02 & 6.70E+02 & 1.65E+03\\

  All-quad generation
        & 1.67E-01 & 7.02E-01 & 1.34E+00 & 4.24E+00 & 1.74E+01 & 4.61E+01 & 1.20E+02\\ 
\hline
\hline
  All-quad to all-hex 
        & 5.64E-02 & 2.48E-01 & 5.13E-01 & 2.42E+00 & 9.34E+00 & 2.52E+01 & 7.97E+01\\
  Extrusion 1
        & 4.99E-01 & 3.58E+00 & 8.60E+00 & 2.11E+01 & 8.58E+01 & 3.10E+02 & 1.63E+03\\ 
\hline
\hline
  IO for smoothing 
        & 2.42e-01 & 4.99E+00 & 4.13E+00 & 1.30E+01 & 5.85E+01 & 1.96E+02 & 1.12E+03\\ 
  Mesh smoothing 
        & 3.58e+00 & 4.12E+01 & 9.95E+01 & 3.99E+02 & 7.26E+02 & 3.19E+03 & 1.10E+03\\ 
  (nodes, $N$)
        &  (1, $N$=1) &(1, $N$=1) &(1, $N$=1) &(2, $N$=1) &(4, $N$=1) &(8, $N$=1) &(24, $N$=1)\\
\hline
\hline
  Extrusion 2
        & 1.01E+00 & 5.36E+00 & 1.08E+01 & 2.80E+01 & 1.10E+02 & 6.72E+02 & 2.12E+03\\
\hline
\hline
  IO for projection
       & 1.55E-01 & 7.71E-01 & 1.62E+00 & 5.13E+00 & 2.19E+01 & 1.62E+02 & 4.16E+02 \\
  Curve-side projection 
       & 4.00E+01 & 2.10E+02 & 1.80E+03 & 1.68E+03 & 4.20E+03 & 3.60E+03 & 7.20E+03 \\
  (nodes, $N$)
       & (1, $N$=2) &(1, $N$=2) &(1, $N$=2) &(2, $N$=2) &(4, $N$=2) &(8, $N$=2) &(600, $N$=7) \\
\hline
\hline
 Total  & 6.53E+01 & 2.70E+02 & 1.91E+03 & 2.04E+03 & 4.27E+03 & 1.09E+03 & 5.10E+04\\

\hline
\end{tabular}
\end{center}
\caption{\label{tb:meshtime}
Timing breakdown of meshing measured in seconds.  Most steps are performed in
serial using Matlab on a workstation (Intel Xeon E5-2630 v3 @2.40GHz) while the
ones performed in parallel are provided with the number of nodes and the
polynomial order $N$.  The smoothing and projection are performed on
OLCF/Summit using 42 cores per node.  Note that the case of ann350k required
two passes through the entire procedure to adjust the edge-collapse tolerance.
}
\end{table*}

\begin{figure*}[t]
\centering
{\setlength{\unitlength}{1.0in}
   \begin{picture}(6.500,3.30)(0,0.1)
      \put(0.20,0.00){\includegraphics[width=6.0in]{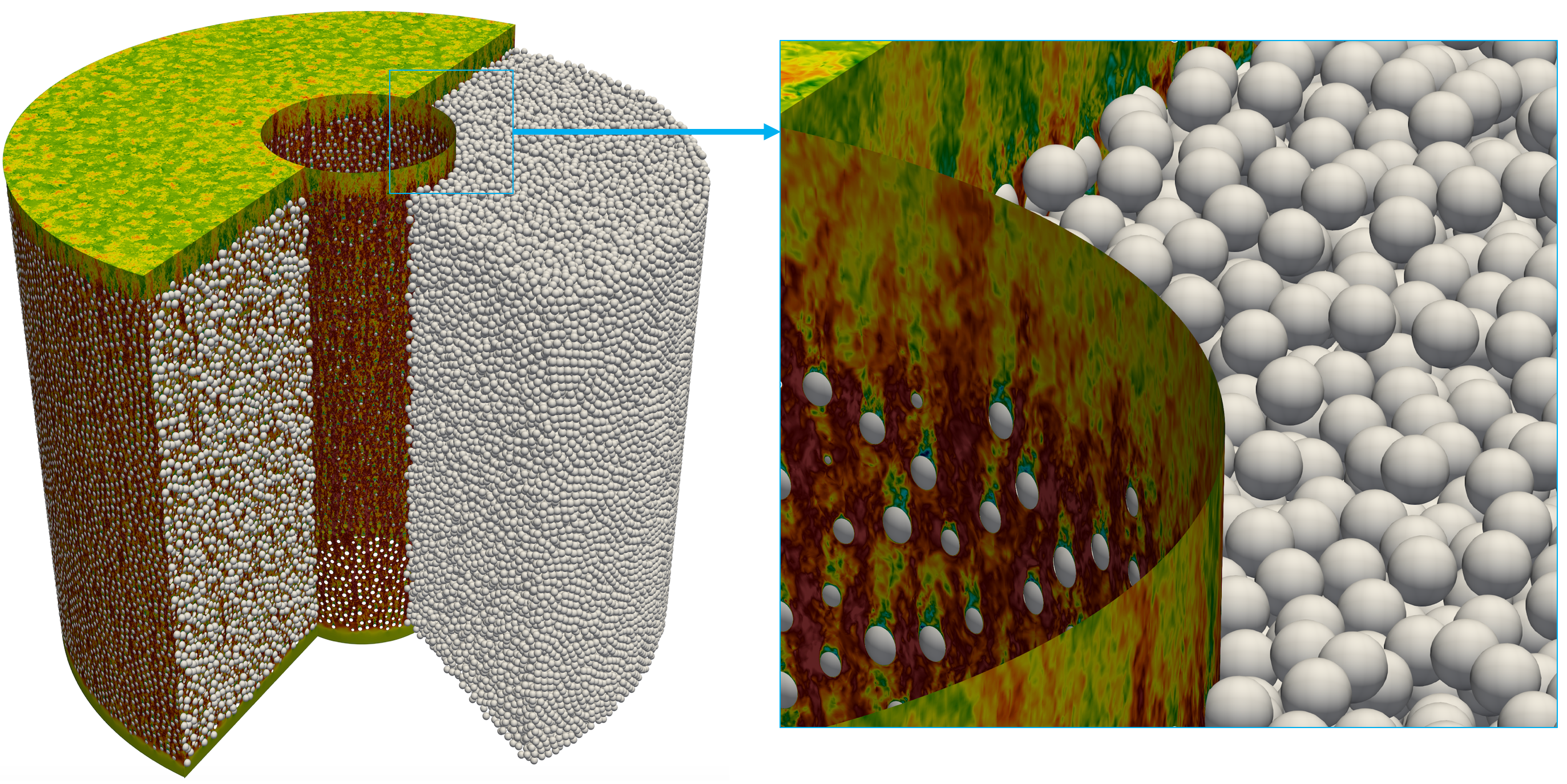}}
   \end{picture}}
\caption{
Turbulent flow in an annular packed bed with $\cN=352625$ spheres meshed with
$E=98,782,067$ spectral elements of order $N=8$ ($n=50$ billion gridpoints).
This NekRS simulation requires 0.233 seconds per step using 27648 V100s on Summit.
The average number of pressure iterations per step is 6.
\label{fig:350k}
}
\end{figure*}

 \begin{table*} [t]
  \footnotesize
  \begin{center}
  \begin{tabular}{|c|c|c|c|c|c|c|c|c|c|c|c|c|}
  \hline
  \multicolumn{13}{|c|}{{\bf Nek5000 performance comparison between hex and tet-to-hex meshes for 146 pebbles}} \\
  \hline
  \hline
   Mesh       & Node & Core &  $E$    &$E$/core &  $N$  &   $n$       & $n$/core & $ v_i$& $p_i$ & CFL  &$t_{step}$  & R     \\
  \hline                                                                                           
   all-hex    & 16   &  672 & 62132  &  92  &  7  & 21311276  & 3.1713e+04  &  2.0  &   5   & 0.92 & 0.2954     & 1     \\
   tet-to-hex & 16   &  672 & 365844 &  544 &  4  & 23414016  & 3.4842e+04  &  1.1  &  17   & 2.11 & 0.9251     & 3.13  \\
  \hline
  \end{tabular}
  \end{center}
\vspace{-4ex}
\caption{\label{tb:peb146} Nek5000 performance comparison of all-hex and
tet-to-hex on Summit (CPU) for $\cN=146$ with $Re=5000$. Simulations
were performed for 200 steps with step size $\dt$=8.00E-04. 
Per-step averages are taken over the last 100 steps.
Here, $v_i$ and $p_i$ represent respective average velocity and pressure
iteration counts, and $t_{step}$ is the average wall clock time, in seconds.
$R$ represents the ratio of all-hex to
tet-to-hex for $t_{step}$. 
Characteristic-based BDF2 with 2 substeps is used for timestepping and
overlapping-Schwarz smoothing with spectral element multigrid preconditioning
with HYPRE AMG for coarse-grid solve is used for
pressure solve. Tolerances for pressure and velocity are  $10^{-4}$ and
$10^{-6}$, respectively.
}
 \end{table*}

\begin{figure*}[!t] \centering
{\setlength{\unitlength}{1.0in}
\begin{picture}(6.500,2.50)(0,0.07)
\put(0.2,0){\includegraphics[width=0.48\textwidth]{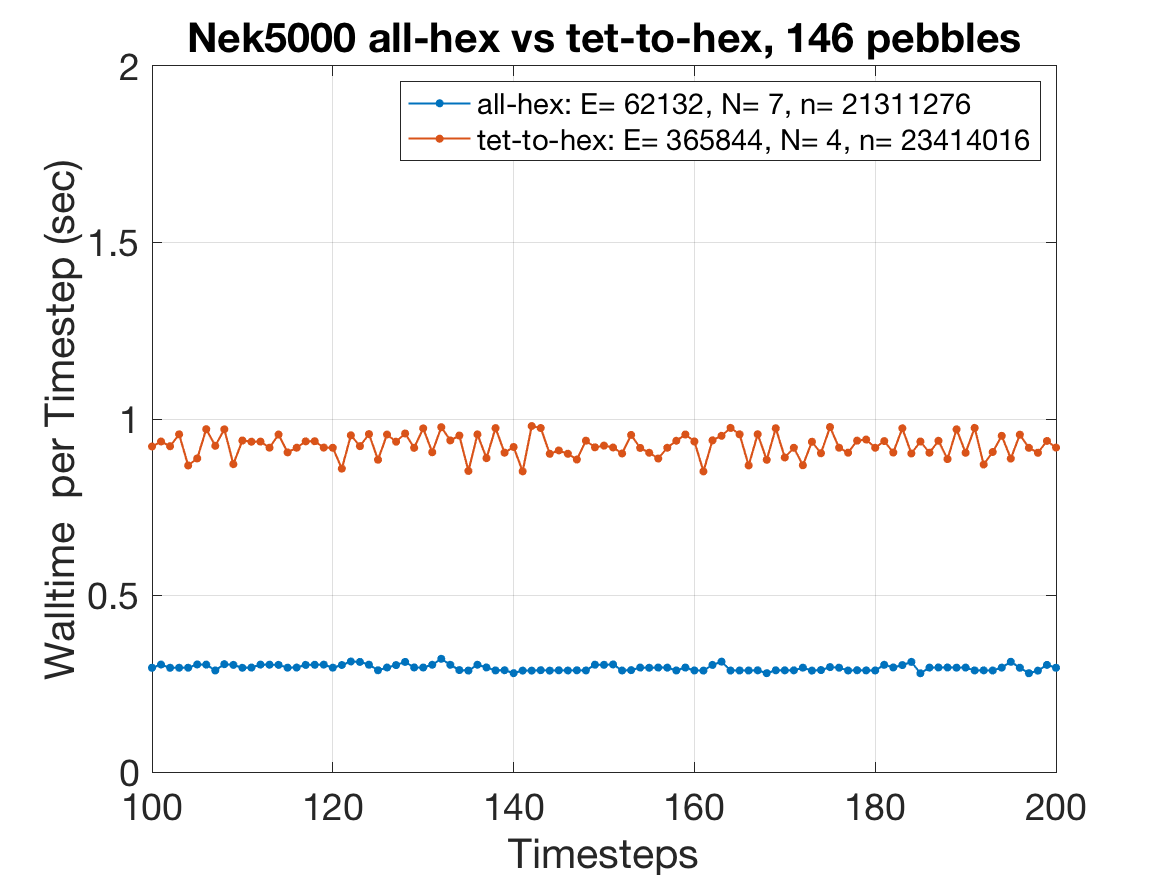}}
\put(3.3,0){\includegraphics[width=0.48\textwidth]{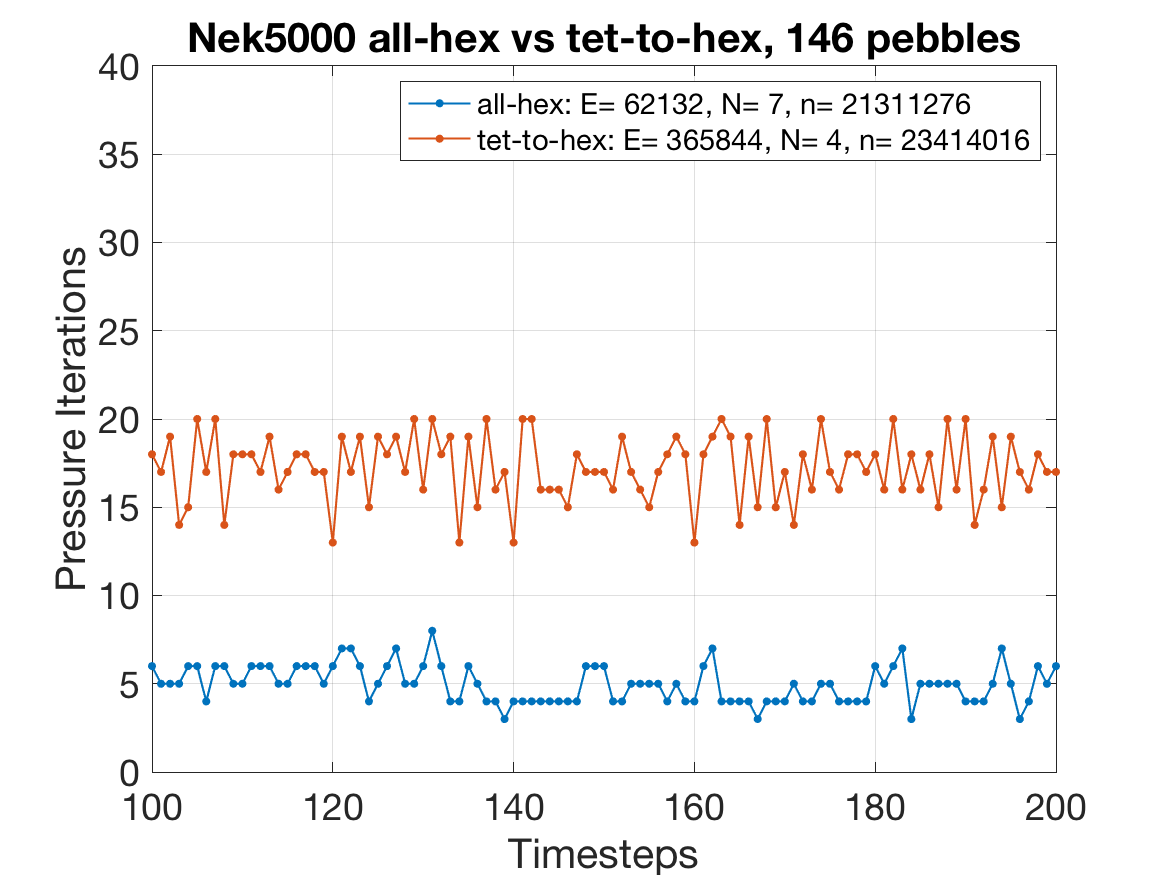}}
\end{picture}}
\caption{\label{fig:peb146}Nek5000 simulation wall time per timestep and
pressure iterations for the case of Table \ref{tb:peb146}.
}
\end{figure*}

\begin{figure*}[!t] \centering
{\setlength{\unitlength}{1.0in}
  \begin{picture}(6.500,2.50)(0,0.07)
     \put(0.7,0){\includegraphics[width=0.38\textwidth]{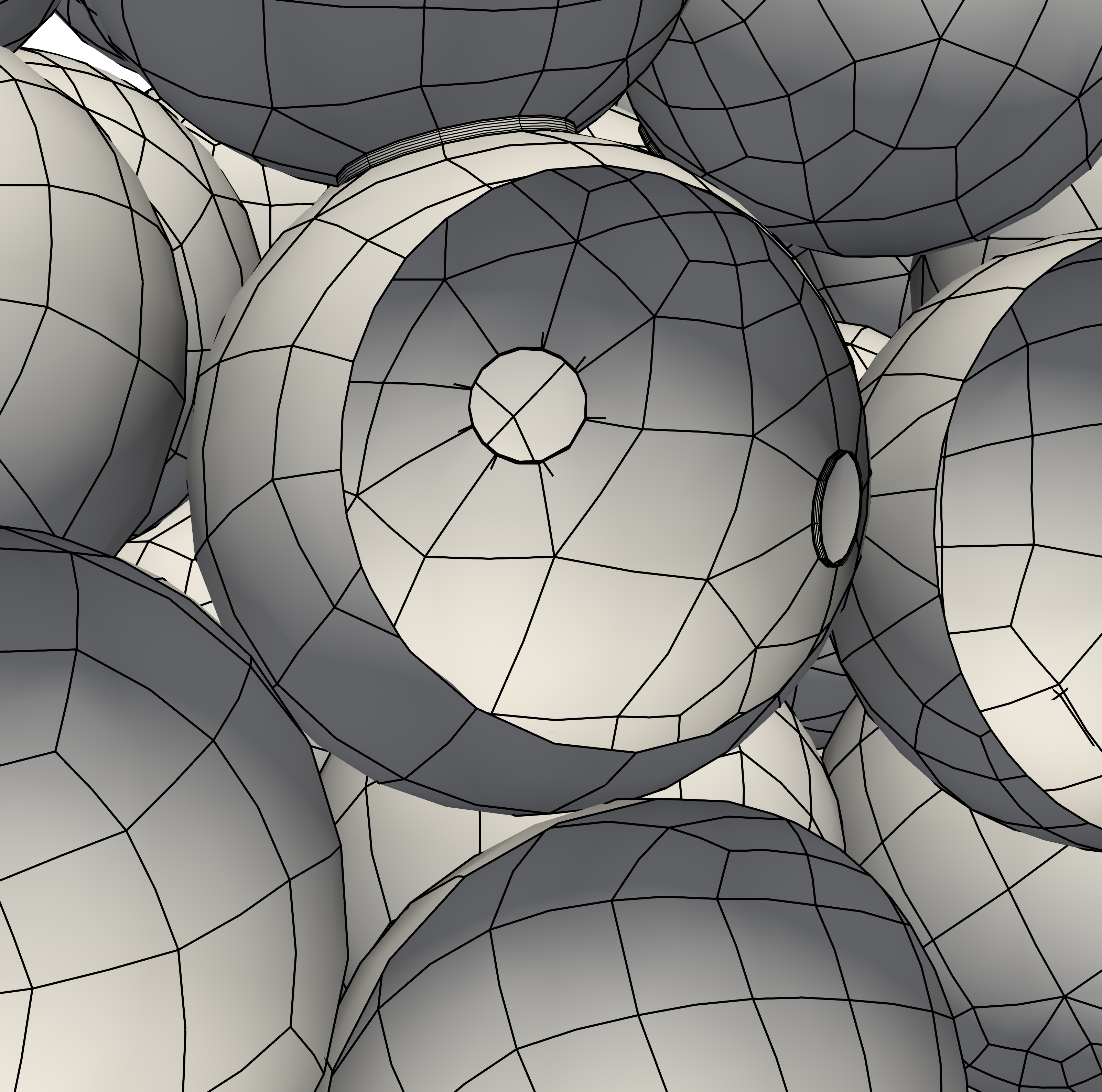}}
     \put(3.4,0){\includegraphics[width=0.38\textwidth]{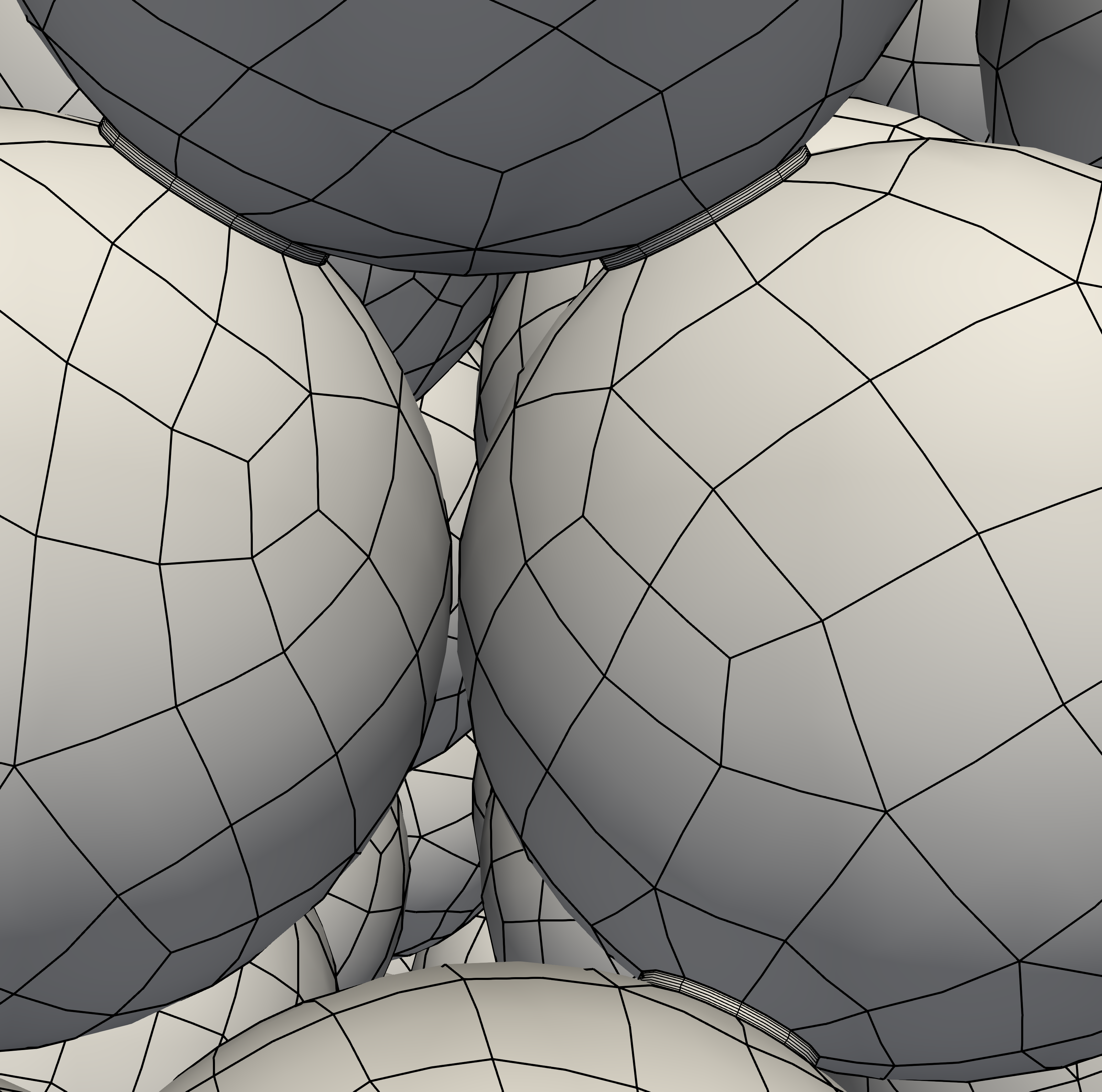}}
 \end{picture}}
  \caption{\label{fig:contact}Prototype mesh for contacting spheres.  Cut-away view on the left.}
  \end{figure*}

\section{Results}

We have applied the algorithm described in Section 2 to the configurations
listed in Table \ref{tb:cases}.  The number of elements per sphere for
the three-layer configuration yields $\approx$300 elements per sphere,
including the elements in the inlet- and exit-flow regions.  Figure
\ref{fig:all_pebbles} shows several of the corresponding sphere configurations
along with axial-flow velocity distributions.  Regions of order and disorder
are evident in the positions of the spheres along the domain wall for the case
$\cN=44252$.  Such ordered packing has a direct influence on the flow
conditions near the domain boundary and is an important consideration for
thermal-hydraulics analysis.  All of the cases shown here were run with NekRS
(the GPU-oriented version of Nek5000) using the NVIDIA V100s on OLCF's Summit
at Oak Ridge National Laboratory.

Table \ref{tb:meshtime} provides a timing breakdown (in seconds) for the
signicant components of the mesher, across the full spectrum of problems
ranging from 62K elements for $\cN$=146 to 99M elements for $\cN$=352K.
Only the 352K case needs to be iterated because of a handful of bad Jacobians
in the mesh using the original edge-collapse tolerance.  
This iteration likely could be avoided with improved smoothing algorithms.
For similar reasons, we perform the final projection and smoothing for the
352K case for the $N=7$ (512 points per hex) configuration, rather than the
standard hex27 used for the other meshes.  We note that the facet generation
and tessellation are bottlenecks because of unavoidable {\em for} loops in
the (interpretive-based) Matlab code.  Rewriting these as $C$-based Mex
files would likely alleviate this bottleneck.   

Regarding mesh optimization, we recall that initial smoothing is applied before
refinement, which means that it is applied to only $\approx 33$M elements for
the $\cN=352$K case of Fig. \ref{fig:350k}.  With 10 outer smoothing passes,
each using 3 Laplacian smoothings followed by 120 optimization steps (requiring
1100 seconds on 1,008 CPU cores of Summit), we find signifcant improvements:
\\[1ex]
\noindent{\bf Before optimization:} \\[-6ex]
\begin{itemize}
    \item 369 elements with negative Jacobians \\[-5ex]
    \item min scaled Jacobian = -5.94 \\[-5ex]
    \item max aspect ratio = 3.63e4 \\[-5ex]
    \item node spacing: (min, max) = (2.32e-5, 7.49e-1) \\[-4ex]
\end{itemize}
\noindent{\bf After optimization:} \\[-6ex]
\begin{itemize}
    \item Valid mesh---all Jacobians positive. \\[-5ex]
    \item min scaled Jacobian = 1.93e-2 \\[-5ex]
    \item max aspect ratio = 3.08e1 \\[-5ex]
    \item node spacing: (min, max) = (4.27e-2, 8.71e-1) \\[-5ex]
\end{itemize}

In Table \ref{tb:peb146} and Fig.  \ref{fig:peb146} we compare the
performance of the all-hex meshing strategy with the tet-to-hex approach
\cite{yuan2020} for the $\cN=146$ case.  At comparable resolution $n$, the
all-hex approach has a lower Courant number (CFL) for the same timestep size
and a lower wall-clock time per timestep because of the reduction in pressure
iteration counts.   The lower CFL implies that the all-hex approach could use a
larger timestep, thereby further reducing simulation costs.

\section{Conclusion and Future Developments}

We have presented an approach to construction of high-quality all-hex
meshes for the interstitial space in dense-packed spheres at relatively
low element counts.  The algorithm uses an $O(\cN)$-complexity Voronoi 
decomposition to decouple the problem into local problems that can 
be meshed independently.   Sliver removal, vertex insertion, automated facet
tessellation, and mesh smoothing are all critical components for generating
production-quality meshes.  All issues regarding boundary conditions and
projection of the final GLL nodal points for the SEM onto the sphere are
addressed.  The success of the algorithm is demonstrated over a broad range
of mesh sizes, from $\cN=146$ to 352K, with the largest case corresponding
to $E=99$ million spectral elements and $n=50.5$ billion grid points.

Overall, the development has satisfied the objective of allowing us to produce
large-scale high-quality meshes suitable for high-order spectral element
simulations of turbulence in packed beds. 
In particular, the 352K case, which corresponds to a full reactor core,
takes only 0.233 seconds per step when running on 4,608 nodes (27,648 V100s),
which corresponds to 1.8 million points per V100.   This configuration would
require only 6 hours to compute a single flow-through time on all of Summit,
implying that parameter studies will be readily tractable on exascale platforms.  
The number of pressure iterations is $\approx$6 per step when using a tuned
version of the NekRS multigrid solver.  Tuning was required because the highly
compressed elements that are squeezed between the nominal sphere contact points
lead to ill-conditioning of the Poisson problem.  

A future development for our mesher will be to replace the flattened spheres at
the contact points with a chamfered bridge of solid material that will join the
spheres and bypass the point-contact singularity, as illustrated in Fig.
\ref{fig:contact}.  Preliminary experience suggests that elimination of the
narrow gap leads to a significant reduction in the condition number of the
discrete pressure-Poisson operator and should further improve runtimes.


\section*{Acknowledgments}
This material is based upon work supported by the U.S. Department of Energy, 
Office of Science and Office of Nuclear Energy, under contract DE-AC02-06CH11357.
%
Programs supporting this research include the DOE 
Exascale Computing Project (17-SC-20-SC), Applied Mathematics Research, and
Nuclear Energy Advanced Modeling and Simulation (NEAMS).
%
  This research used resources of the Oak Ridge Leadership
  Computing Facility at Oak Ridge National Laboratory, which is supported by the
  Office of Science of the U.S. Department of Energy under Contract DE-AC05-00OR22725.


\bibliography{imr.bbl}

\begin{thebibliography}{10}
\newcommand{\enquote}[1]{``#1''}
\expandafter\ifx\csname url\endcsname\relax
  \def\url#1{\texttt{#1}}\fi
\expandafter\ifx\csname urlprefix\endcsname\relax\def\urlprefix{URL }\fi

\bibitem{packedbed06}
Kolev N.
\newblock \emph{Packed bed columns: for absorption, desorption, rectification
  and direct heat transfer}.
\newblock Elsevier, 2006

\bibitem{merzari2020a}
Merzari E., Yuan H., Min M., Shaver D., Rahaman R., Shriwise P., Romano P.,
  Talamo A., Lan Y., Gaston D., Martineau R., Fischer P., Hassan Y.
\newblock \enquote{Cardinal: A lower length-scale multiphysics simulator for
  pebble bed reactors.}
\newblock \emph{Nucl. Tech., American Nuclear Society}, 2020

\bibitem{pat84}
Patera A.
\newblock \enquote{A spectral element method for fluid dynamics : laminar flow
  in a channel expansion.}
\newblock \emph{J. Comp. Phys.}, vol.~54, 468--488, 1984

\bibitem{stephenson1992}
Stephenson M.B., Canann S.A., Blacker T.D.
\newblock \enquote{Plastering: a new approach to automated, 3{D} hexahedral
  mesh generation--Progress Report {I}.}
\newblock Tech. Rep. 89-2192, Sandia National Labs., Albuquerque, 1992

\bibitem{cass1996}
Cass R., Benzley S., Meyers R., Blacker T.
\newblock \enquote{Generalized {3-D} paving: an automated quadrilateral surface
  mesh generation algorithm.}
\newblock \emph{Int. J. for Num. Meth. in Eng.}, vol.~39, no.~9, 1475--1489,
  1996

\bibitem{geode}
Leland R., Melander D., Meyers R., Mitchell S., Tautges T.
\newblock \enquote{The Geode Algorithm: Combining Hex/Tet Plastering, Dicing
  and Transition Elements for Automatic, All-Hex Mesh Generation.}
\newblock \emph{IMR}, pp. 515--521. 1998

\bibitem{mcdill2004}
McDill J., Carmona~Garcia A.
\newblock \enquote{Tet-to-Hex Conversion for Finite Element Analysis.}
\newblock \emph{AIP Conference Proceedings}, vol. 712, pp. 2210--2215. American
  Institute of Physics, 2004

\bibitem{yuan2020}
Yuan H., Yildiz M., Merzari E., Yu Y., Obabko A., Botha G., Busco G., Hassan
  Y., Nguyen D.
\newblock \enquote{Spectral element applications in complex nuclear reactor
  geometries: Tet-to-hex meshing.}
\newblock \emph{Nuclear Engineering and Design}, vol. 357, 110422, 2020

\bibitem{fischer20a}
Fischer P., Min M., Rathnayake T., Dutta S., Kolev T., Dobrev V., Camier J.,
  Kronbichler M., Warburton T., Swirydowicz K., Brown J.
\newblock \enquote{Scalability of High-Performance {PDE} Solvers.}
\newblock \emph{IJHPCA}, vol. 34, 5, 562--586, 2020

\bibitem{cruz_patera_95}
Cruz M.E., Patera A.T.
\newblock \enquote{A parallel {Monte Carlo} finite-element procedure for the
  analyis of multicomponent random media.}
\newblock \emph{J. Num. Meth. Eng.}, vol.~38, 1087--1121, 1995

\bibitem{cruz_ghaddar_patera_95}
Cruz M.E., Ghaddar C.K., Patera A.T.
\newblock \enquote{A variational-bound nip-element method for geometrically
  stiff problems; application to thermal composites and porous media.}
\newblock \emph{Proc: Math. Phys. Sci}, vol. 449, 93--122, 1995

\bibitem{sheffer_voronoi_99}
Sheffer A., Etzion M., Rappoport A., Bercovier M.
\newblock \enquote{Hexahedral Mesh Generation using the Embedded Voronoi
  Graph.}
\newblock \emph{Engineering with Computers}, vol.~15, no.~3, 248--262, 1999

\bibitem{levy2010}
Yan D.M., Wang W., L{\'e}vy B., Liu Y.
\newblock \enquote{Efficient Computation of {3D} Clipped Voronoi Diagram.}
\newblock pp. 269--282, 2010

\bibitem{yassin}
Nguyen T., Kappes E., King S., Hassan Y., Ugaz V.
\newblock \enquote{Time-resolved PIV measurements in a low-aspect ratio
  facility of randomly packed spheres and flow analysis using modal
  decomposition.}
\newblock \emph{Experiments in Fluids}, vol.~59, no.~8, 1--29, 2018

\bibitem{mittal19a}
Mittal K., Fischer P.
\newblock \enquote{Mesh Smoothing for the Spectral Element Method.}
\newblock \emph{J. Sci. Comput.}, vol.~78, no.~2, 1152--1173, 2019

\bibitem{knupp1}
Knupp P.
\newblock \enquote{Introducing the target-matrix paradigm for mesh optimization
  via node-movement.}
\newblock \emph{Engineering with Computers}, vol.~28, no.~4, 419--429, 2012

\bibitem{dfm02}
Deville M., Fischer P., Mund E.
\newblock \emph{High-order methods for incompressible fluid flow}.
\newblock Cambridge University Press, Cambridge, 2002

\bibitem{stfi73}
Strang G., Fix G.
\newblock \emph{An Analysis of the Finite Element Method}.
\newblock Prentice-Hall Series in Automatic Computation. Prentice-Hall,
  Englewood Cliffs, NJ, 1973

\bibitem{knupp2}
Knupp P.M.
\newblock \enquote{Hexahedral Mesh Untangling \& Algebraic Mesh Quality
  Metrics.}
\newblock \emph{IMR}, pp. 173--183. Citeseer, 2000

\bibitem{knupp3}
Knupp P.M.
\newblock \enquote{A method for hexahedral mesh shape optimization.}
\newblock \emph{Int. J. Num. Meth. in Eng.}, vol.~58, no.~2, 319--332, 2003

\bibitem{edge1}
Livesu M., Sheffer A., Vining N., Tarini M.
\newblock \enquote{Practical hex-mesh optimization via edge-cone
  rectification.}
\newblock \emph{ACM Transactions on Graphics (TOG)}, vol.~34, no.~4, 1--11,
  2015

\bibitem{edge2}
Xu K., Gao X., Chen G.
\newblock \enquote{Hexahedral mesh quality improvement via edge-angle
  optimization.}
\newblock \emph{Computers \& Graphics}, vol.~70, 17--27, 2018

\end{thebibliography}

\bigskip
     The submitted manuscript has been created by UChicago Argonne, LLC, Operator of Argonne National Laboratory (“Argonne”). Argonne, a U.S. Department of Energy Office of Science laboratory, is operated under Contract No. DE-AC02-06CH11357. The U.S. Government retains for itself, and others acting on its behalf, a paid-up nonexclusive, irrevocable worldwide license in said article to reproduce, prepare derivative works, distribute copies to the public, and perform publicly and display publicly, by or on behalf of the Government. The Department of Energy will provide public access to these results of federally sponsored research in accordance with the DOE Public Access Plan. http://energy.gov/downloads/doe-public-access-plan

\end{document}